\documentstyle[amsmath,amssymb,epsfig,12pt,arydshln,psfrag,graphicx]{article}

\begin{document}
\setlength{\parskip}{0.45cm}
\setlength{\baselineskip}{0.75cm}
%
%
%
\begin{titlepage}
\setlength{\parskip}{0.25cm}
\setlength{\baselineskip}{0.25cm}
\begin{flushright}
DO-TH 2003/03\\
\vspace{0.2cm}
astro--ph/0303252\\
\vspace{0.2cm}
March 2003
\end{flushright}
\vspace{1.0cm}
\begin{center}
\LARGE
{\bf Cosmic UHE Neutrino Signatures}
\vspace{1.5cm}

\large
K.\ Giesel, J.-H.\ Jureit, E.\ Reya\\
\vspace{1.0cm}

\normalsize
{\it Universit\"{a}t Dortmund, Institut f\"{u}r Physik,}\\
{\it D-44221 Dortmund, Germany} \\
\vspace{0.5cm}

\vspace{1.5cm}
\end{center}

\begin{abstract}
\noindent Utilizing the unique and reliable
ultrasmall--$x$ predictions of the dynamical
(radiative) parton model, nominal event rates
and their detailed energy dependence caused by
a variety of cosmic UHE neutrino fluxes are 
calculated and analyzed.  In addition, maximal
Regge--model inspired small--$x$ structure 
functions are employed for obtaining optimal
rates which do 
not necessarily require `new'
physics interpretations.  Upward $\mu^+ +\mu^-$
event rates are estimated by taking into account
total and nadir--angle dependent regeneration
effects due to neutral current interactions.
For exploring extragalactic neutrino sources
at highest energies
(\raisebox{-0.1cm}{$\stackrel{>}{\sim}$} 
$10^8$ GeV)
with modern (future) ground--level telescopes,
we analyze horizontal air shower event rates
and shower
events caused by Earth--skimming tau--neutrinos,
in particular their detailed shower-- and
cosmic neutrino--energy dependence.  As an
illustration of `new physics' implications we
estimate the relevant horizontal air shower
event rates due to spin--2 Kaluza--Klein
`graviton' exchanges in neutral current 
neutrino--quark and neutrino--gluon interactions
at low TeV--scales. 
\end{abstract}
\end{titlepage}


\renewcommand{\theequation}{\arabic{section}.\arabic{equation}}
\section{Introduction}
Detection of cosmic ultrahigh--energy (UHE)
neutrinos with energies above $10^{16}$ eV is
one of the important challenges of cosmic ray
detectors in order to probe the faintest regions
of the Universe that are otherwise shielded
from us by large amounts of matter.  Their
observation will probe particle (possibly
`new') physics as well as astrophysics phenomena
such as galaxy formation.  The sources of UHE
neutrinos range, however, from the well 
established to the highly speculative 
\cite{ref1,ref2,ref3,ref4,ref5}.  The well
measured cosmic rays up to the GZK cutoff
\cite{ref6} at around $5\times 10^{19}$ eV,
where they necessarily interact with the 2.7 K
cosmic microwave background through $p\gamma
\to n\pi^+$, produce the  `guaranteed' (cosmogenic)
flux of `Greisen neutrinos' when the pions decay
\cite{ref6,ref7}.  In addition, far larger
neutrino fluxes are predicted in models of
active galactic nuclei (AGN) 
\cite{ref8,ref9,ref10}, gamma ray bursts
(GRB) \cite{ref11}, decays of exotic heavy
particles of generic top--down or topological
defects (TD) \cite{ref12,ref13,ref14,ref15,ref16}
and $Z$--bursts \cite{ref17,ref18}.  
Representative fluxes of some hypothesized
sources are displayed in Fig.\ 1 which we
shall use for all our subsequent calculations.
For illustration the steeply falling background
atmospheric (ATM) neutrino flux
\cite{ref19,ref20,ref21} is shown as well which
originates from (anti)neutrinos produced by
cosmic ray interactions in the Earth's atmosphere.

Besides these violently different model
expectations for (anti)neutrino fluxes, there
are further uncertainties when calculating
event rates for neutrino telescopes due to 
the sensitivity of $\stackrel{(-)}{\nu}\!\!\!N$
cross sections to the parton distributions in
the yet unmeasured ultrasmall Bjorken--$x$
region, $x<10^{-5}$, not accessible at the 
weak scale $M_W$ by deep inelastic scattering
(DIS) experiments.  At the highest 
(anti)neutrino energies $E_{\nu}$ shown
in Fig.\ 1, contributions from the region
around 
$x\simeq M_W^2/2M_N E_{\nu}\simeq 10^{-8}$
to $10^{-9}$
become non--negligible and therefore known 
structure functions and parton distributions
at $x$ \raisebox{-0.1cm}{$\stackrel{>}{\sim}$}
$10^{-5}$
have to be extrapolated to $x<10^{-5}$ as 
soon as $E_{\nu}$ \raisebox{-0.1cm}{$\stackrel
{>}{\sim}$} $10^8$ GeV.  (Here one commonly
assumes that the DIS small--$x$ measurements 
at $x$ \raisebox{-0.1cm}{$\stackrel{>}{\sim}$}
$10^{-5}$ and moderate momentum scales $Q^2$
can be safely evoluted to the relevant scale
$Q^2=M_W^2$ by standard QCD renormalization
group (RG) techniques.)  Such extensive  
extrapolations are performed either of fits
to existing data at $x$ 
\raisebox{-0.1cm}{$\stackrel{>}{\sim}$}
$10^{-5}$ using specific, possibly arbitrary
and unreliable, assumptions (e.g.\ various
fixed power behaviors in $x$ of structure
functions as $x\to 0$) 
\cite{ref20,ref21,ref22,ref23,ref24,ref25,ref26}
or by using \cite{ref27,ref28} the QCD inspired
dynamical (radiative) parton model which proved
to provide reliable high energy predictions
\cite{ref29,ref30,ref31,ref32} in the past
\cite{ref33,ref34}.  Within this latter
approach the entire partonic (gluons and
(anti)quarks) structure at small fractional
momenta $x$ \raisebox{-0.1cm}{$\stackrel{<}
{\sim}$} $10^{-2}$ can be understood and
calculated via RG evolutions from first principles,
i.e.\ QCD dynamics, independently of any
free (fit) parameter in the small--$x$ region
due to valence--like gluon and sea input 
densities at some low momentum scale
$Q_0\simeq$ 0.5 -- 0.6 GeV 
\cite{ref30,ref31,ref32}. Having successfully
predicted the small--$x$ behavior of structure
functions between $x=10^{-2}$ to $10^{-5}$ prior
to experiments, it is not unreasonable to 
expect the unambiguous dynamical results of
the radiative parton model between 
$x=10^{-5}$ and $10^{-8}$ or $10^{-9}$ to be
reliable as well.  These ultrasmall--$x$ 
predictions are furthermore perturbatively
stable and unique at the relevant momentum
scales $Q^2\simeq M_W^2$ \cite{ref32}.  
Moreover, consistent BFKL model resummations
of subleading $\ln \frac{1}{x}$ contributions
in leading order QCD yield remarkably similar
results \cite{ref35} even at highest neutrino
energies of $10^{12}$ GeV.

A detailed analysis \cite{ref36} has shown
that within the radiative parton model all
relevant (anti)neutrino--nucleon cross sections
can be calculated with an uncertainty of
typically about $\pm 20\%$ at highest neutrino
energies of $10^{12}$ GeV.  We shall adopt 
this approach for all our subsequent 
calculations, in particular the canonical 
GRV98 parton distributions \cite{ref32} in
next--to--leading order QCD will serve as
our appropriate nominal benchmark set of 
densities.  Notice that cross sections obtained
from the fitted CTEQ3--DIS parametrizations
\cite{ref37} at 
$x$ \raisebox{-0.1cm}{$\stackrel{>}{\sim}$} 
$10^{-5}$ with their assumed fixed--power 
extrapolation to $x<10^{-5}$, as used in
\cite{ref20}, accidentally coincide practically
with the ones derived from the ultrasmall--$x$
dynamical predictions of the radiative parton 
model \cite{ref36}; only at highest neutrino 
energies of $10^{12}$ GeV, i.e.\ $x\simeq 
10^{-8}$ to $10^{-9}$, they are about 10\% 
larger than the radiatively generated ones.
Actual calculations of event rates could 
therefore also be performed with the CTEQ3--DIS
densities which are easier to use since heavy
quarks ($c,\, b,\, t$) are effectively (but 
less adequately) treated as massless intrinsic 
partons, in contrast to the more appropriate
but also more cumbersome explicit calculation
of massive heavy quark contributions required
when using the `fixed flavor' GRV98 densities
\cite{ref32}. (A comparison of the two approaches
can be found in \cite{ref36}.)  Similar remarks
hold true for the CTEQ4--DIS parametrizations
\cite{ref38}, used in \cite{ref21}, which
underestimate, relatively to GRV98, the neutrino
cross sections by about 20\% at $E_{\nu}\simeq
10^{10}$ -- $10^{12}$ GeV.  

The detection of UHE cosmic neutrinos is, however,
extremely difficult.  For energies below $10^8$
GeV it is widely believed that one of the most
appropriate techniques for neutrino detection 
consists of detecting the \v{C}erenkov
light from muons or showers produced by CC and
NC neutrino interactions of mainly upward--going
neutrinos in large--volume underground water or
ice detectors such as AMANDA/IceCube, NESTOR
and ANTARES \cite{ref2,ref39}.  However, above
40 TeV the Earth's diameter exceeds the interaction
length of neutrinos and their shadowing in the
Earth rapidly increases above 100 TeV 
\cite{ref20,ref21} which severely restricts rates
in underground detectors which are bounded by
detection volumes of at most 1 km$^3$.
Eventually it becomes beneficial to look for 
events induced by downward--going and 
(quasi)horizontal neutrinos, provided of course
that downward--going events produced by 
interactions within the instrumented underground
detector volume can be efficiently observed. At
energies above $10^8$ GeV, where the 
(anti)neutrino interaction length is below about
10$^3$ km water equivalent in rock, upward--going
neutrinos are blocked by Earth and thus 
under--water and under--ice km--scale detectors
become ineffective due to the opaqueness of Earth
to upward neutrinos.  Therefore large--area 
ground arrays or surface fluorescence telescopes
such as
AGASA, the HiRes detector, the Pierre Auger
Observatory and the Telescope Array 
\cite{ref26,ref28,ref40,ref41}, for which the 
interaction medium is not the Earth but the 
atmosphere, could become instrumental in 
exploring the whole spectrum of cosmic neutrino
fluxes up to highest neutrino energies of about
$10^{21}$ eV shown in Fig.\ 1. In particular,
a novel alternative method for detecting UHE
(anti)neutrinos with $E_{\nu}>10^8$ GeV has
been recently suggested via Earth--skimming
neutrinos \cite{ref42,ref43,ref25}, dominantly 
$\nu_{\tau}$ travelling at large nadir angles 
producing `double bang' events in the atmosphere 
when converting to tau--leptons exiting the 
Earth's surface which, when decaying, produce the 
second shower bang. 

When upward--going UHE neutrinos penetrate 
through the Earth, they undergo attenuation
(absorption) due to charged and neutral current
interactions as well as regeneration 
\cite{ref44,ref45} due to the neutral current
interactions (which shift their energy rather
than absorbing them) at high energies.  The
NC shifts the neutrino energies to lower
energies and, in addition, regeneration 
populates the lower energy part of the flux
spectra shown in Fig.\ 1.  In Sect.\ 2 we 
shall perform detailed calculations of these
regeneration effects \cite{ref35} not only for
our representative fluxes in Fig.\ 1 but in 
particular also for the expected event rates
for modern underground detectors.  Depending
on the shape of the neutrino flux, these
effects can be substantial, i.e.ß increase
the non--regenerated event rates on the 
average by about 20\%.  For comparison we
therefore have also to recalculate, for our
representative fluxes in Fig.\ 1, the upgoing
event rates where regeneration effects have
not been taken into account when calculating
the shadow factor \cite{ref20,ref21,ref23}.
In addition we show in detail what high--energy
bins of the original fluxes in Fig.\ 1 remain
experimentally accessible after their depletion
and regeneration due to NC interactions at
highest neutrino energies.  Furthermore, 
downward events initiated by UHE neutrinos
will be also calculated using recent 
Regge--inspired  power--like small--$x$
extrapolations in order to learn about possible
upper bounds of event rates which may be
accommodated by `conventional' standard model
approaches and which do not necessarily 
require  `new' physics interpretations. Along
similar lines we study in Sect.\ 3 event rates
of quasi--horizontal air showers to be detected 
by large area ground arrays and surface air
fluorescence telescopes, as well as of 
Earth--skimming UHE $\nu_{\tau}
$'s which will
allow to test cosmic neutrino flux models at
highest energies, $E_{\nu}$ 
\raisebox{-0.1cm}{$\stackrel{>}{\sim}$} $10^8$
GeV.  We shall in particular be interested to
what extent shower rates as measured in certain
bins of $E_{\rm sh}$ or $E_{\tau}$ will be 
able to explore and delineate specific features
of the energy profile of the various predicted
incoming cosmic neutrino fluxes $\Phi(E_{\nu})$ 
in Fig.\ 1.  Finally, as an illustration of
possible implications of `new' physics, we present 
in Sect.\ 4 highly speculative expectations of
string theories with large  `extra dimensions'
for the relevant quasi--horizontal air shower
event rates.  Section~5 summarizes our results.

\renewcommand{\theequation}{\arabic{section}.\arabic{equation}}
\section{Event Rates for Underground 
\v{C}erenkov--Detectors}
The upward-muon event rate depends on the
$\stackrel{(-)}{\nu}\!\!\!_{\mu}N$ cross section
through the interaction length $\lambda_{\rm int}$
that governs the attenuation of the original
neutrino flux due to interactions in the Earth
as described by the shadow factor $S$ as well
as through the probability $P_{\mu}$ that the
neutrino converts to a muon energetic enough
to arrive at the detector with energy $E_{\mu}$
larger than the assumed threshold energy 
$E_{\mu}^{\rm{min}}$.  The total (nadir angle
integrated) upward--muon event rate per second
in a detector with an energy dependent area
$A(E_{\mu})$ is then given by \cite{ref20,ref21}
\begin{equation}
\rm{rate} = 2\pi \int_{E_{\mu}^{\rm min}} dE_{\nu}
  \int_{0}^{1-E_{\mu}^{\rm min}/E_{\nu}} dy\,
   A(E_{\mu})P_{\mu}(E_{\nu},y;\, E_{\mu}^{\rm min})
    S(E_{\nu})\, \frac{d\Phi_{\nu}}{dE_{\nu}}
\end{equation}
where the factor of $2\pi$ is the effective
solid angle for upward muons, 
$E_{\mu}=(1-y)E_{\nu}$ and $\Phi_{\nu}(E_{\nu})$
refers to the original cosmic neutrino fluxes
shown in Fig.\ 1 incident on the surface of 
the Earth (i.e.\ $\Phi_{\nu,\bar{\nu}}=\Phi/4$)
\underline{before} they undergo attenuation and
regeneration.  Unless otherwise stated, the 
energy integration is always extended to the 
highest energy of $10^{12}$ GeV in Fig.\ 1.
Furthermore,
\begin{equation}
P_{\mu}(E_{\nu},y;\, E_{\mu}^{\rm min}) =
 N_A\,R(E_{\mu},\, E_{\mu}^{\rm min}) \,
  \frac{d\sigma_{CC}^{\nu N}(E_{\nu},y)}{dy}
\end{equation}
with $N_A=6.022\times 10^{23}\,g^{-1}$ and where 
all relevant expressions for calculating CC 
and NC $\stackrel{(-)}{\nu}\!\!\!N$ cross sections
in the  `fixed flavor' and   `variable flavor'
scheme can be found in \cite{ref36}.  It may 
be helpful to notice that, for an 
energy--independent effective detector area
$A_{\rm eff}$, (2.1) reduces to the more
common expression \cite{ref20,ref21}
\begin{equation}
\rm{rate} = 2\pi A_{\rm eff} 
  \int_{E_{\mu}^{\rm min}} dE_{\nu}\, P_{\mu}
   (E_{\nu};\, E_{\mu}^{\rm min})\,S(E_{\nu})\,
    \frac{d\Phi_{\nu}}{dE_{\nu}}
\end{equation}
with
\begin{equation}
P_{\mu}(E_{\nu};\, E_{\mu}^{\rm min}) =
 N_A\, \sigma_{CC}^{\nu N}(E_{\nu})\,
  \langle R \,(E_{\nu};\, E_{\mu}^{\rm min})\rangle
\end{equation}
and the average range $\langle R\rangle$ of a
muon in rock is given by \cite{ref20}
\begin{equation}
\langle R\, (E_\nu;\, E_{\mu}^{\rm min})\rangle 
 = \frac{1}{\sigma_{CC}^{\nu N}(E_{\nu})}\,
  \int_{0}^{1-E_{\mu}^{\rm min}/E_{\nu}} dy\, 
   R\left( (1-y)E_{\nu},\, 
     E_{\mu}^{\rm min}\right) \frac
     {d\sigma_{CC}^{\nu N}(E_{\nu},y)}{dy}\, .
\end{equation}
The range $R$ of an energetic muon in (2.2)
and (2.5) follows from the energy--loss relation
\cite{ref46}
\begin{equation}
-dE_{\mu}/dX = \alpha_{\mu}(E_{\mu})
  +\beta_{\mu}(E_{\mu})\, E_{\mu}
\end{equation}
with $X$ being the thickness of matter traversed
by the muon in units of $g/{\rm cm}^2=$ cm we.  Despite
the very weak energy dependence \cite{ref46} of 
the ionization loss $\alpha_{\mu}(E_{\mu})$ we use
\cite{ref20,ref46} $\alpha_{\mu}=2.0\times 10^{-3}$ 
GeV (cm we)$^{-1}$ since the effect of $\alpha_{\mu}$ 
is negligible for high energies.  Assuming the
fractional energy loss $\beta_{\mu}(E_{\mu})$
to be energy independent as well, for example
\cite{ref20,ref21,ref46} $\beta_{\mu}=3.9\times
10^{-6}$ (cm we)$^{-1}$ or \cite{ref25,ref46}
$\beta_{\mu}=6.0\times 10^{-6}$ (cm we)$^{-1}$,
the relation (2.6) can be integrated analytically,
\begin{equation}
R(E_{\mu},\, E_{\mu}^{\rm min})\equiv
  X(E_{\mu}^{\rm min})-X(E_{\mu}) = 
   \frac{1}{\beta_{\mu}}\, \ln\, 
    \frac{\alpha_{\mu}+\beta_{\mu}E_{\mu}}
     {\alpha_{\mu}+\beta_{\mu}E_{\mu}^{\rm min}}
      \, .
\end{equation}
Alternatively, using a QED--oriented energy
dependence \cite{ref47}
\begin{eqnarray}
\alpha_{\mu}(E_{\mu}) & = & \left[ 2.033+0.077\,
  \ln(E_{\mu}/\rm{GeV})\right]\times 10^{-3}\,\,
   \rm{GeV\,\,(cm\,\,we)}^{-1}\nonumber\\
\beta_{\mu} (E_{\mu}) & = & \left[ 2.229+0.2\,
  \ln(E_{\mu}/\rm{GeV})\right]\times 10^{-6}\,\,
   \rm{(cm\,\,we)}^{-1}\, ,
\end{eqnarray}
(2.6) has to be integrated numerically. In Fig.\ 2
we show the average muon ranges resulting from 
these three different choices and compare them
with the Monte Carlo result of Lipari and Stanev
\cite{ref48} where regeneration has been taken into
account and which has been presented only up to
$10^9$ GeV.  Since this latter Monte Carlo result 
agrees best with our result using a constant
$\alpha_{\mu}$ and $\beta_{\mu}=6.0\times 10^{-6}$
\mbox{(cm we)$^{-1}$}, this choice appears to be best suited for 
extrapolations beyond $10^9$ GeV and we therefore
will utilize it for our subsequent calculations.
Needless to say that this conclusion is 
independent of the specific choice of parton
distributions since obviously $\langle R\rangle$
in (2.5) does practically not depend on them.
(If instead the numerical range of Fig.\ 2 were
used, the resulting event rates would be about
30\% larger.)

Next we turn to the shadow factor in (2.1) 
describing the attenuation (due to 
$\sigma_{\rm tot}= \sigma_{CC}+\sigma_{NC}$)
and regeneration (due to $\sigma_{NC}$) of 
(anti)neutrinos when penetrating through the
Earth.  Both effects are summarized in the 
transport equation \cite{ref44,ref45} for the
neutrino flux $\Phi_{\nu}(E,X)$ at `depth' $X$
\begin{equation}
\frac{d\Phi_{\nu}(E_{\nu},X)}{dX} = -
 \frac{1}{\lambda_{\rm int}(E_{\nu})}\, \Phi_{\nu}
  (E_{\nu},X)+N_A\int_0^1\frac{dy}{1-y}\,\,
   \frac{d\sigma^{\nu N}_{NC}(E_y,y)}{dy}\,
    \Phi_{\nu}(E_y,X)
\end{equation}
where the neutrino interaction length is defined 
by $\lambda_{\rm int}=1/(N_A
\sigma_{\rm tot}^{\nu N})$, the fractional energy
loss by $E_y=E_{\nu}/(1-y)$ and $\Phi_{\nu}
(E_{\nu},0)=\Phi_{\nu}$ where $\Phi_{\nu}=
\Phi_{\nu}(E_{\nu})$ is related to the initial
unmodified fluxes as shown in Fig.\ 1.  It turns
out to be convenient to solve this equation
iteratively via the ansatz \cite{ref35}
\begin{equation}
\Phi_{\nu}(E_{\nu},X)/\Phi_{\nu}(E_{\nu}) =
 \exp \left[ -N_A\sigma_{\rm tot}^{\nu N}
  (E_{\nu})X\right] \Psi_{\nu}(E_{\nu},X)\, ,
\end{equation}
i.e.,
\begin{eqnarray}
\frac{d\Psi_{\nu}(E_{\nu},X)}{dX} & = & 
 N_A \int_0^1 \frac{dy}{1-y}\, 
 \frac{\Phi_{\nu}(E_y)}{\Phi_{\nu}(E_{\nu})}
  \nonumber\\
& &  \exp
   \left\{\!-N_A\left[ \sigma_{\rm tot}^{\nu N}
    (E_y)\!-\sigma_{\rm tot}^{\nu N}(E_{\nu})
     \right]\!X \right\} 
      \frac{d\sigma_{NC}^{\nu N}(E_y,y)}{dy}\,
       \Psi_{\nu}(E_y,X).
\end{eqnarray}
(It should be noticed that this solution can be
found even more efficiently by using \cite{ref45},
instead of the r.h.s.\ of (2.10), $\exp\left[
-X/\Lambda_{\nu}(E_{\nu},X)\right]$ with an
effective absorption length $\Lambda_{\nu}
(E_{\nu},X)\equiv \lambda_{\rm int}/
\left[1-Z_{\nu}(E_{\nu},X)\right]$.) 
According to (2.10), the total shadow factor
is given by
\begin{equation}
S(E_{\nu})=\frac{1}{2\pi}\int_0^{2\pi}d\varphi
 \int_0^{\pi/2}d\theta\sin\theta\exp
  \left[ -N_A\, \sigma_{\rm tot}^{\nu N}
   (E_{\nu})X(\theta)\right] \Psi_{\nu}
     \left( E_{\nu},\, X(\theta)\right)
\end{equation}
where the $d\varphi$ integration is trivial for our 
isotropic (anti)neutrino fluxes.  The amount 
of material encountered by an upward--going
neutrino in its passage through the Earth is
given by the column depth $X(\theta)$ which
depends upon the nadir angle $\theta$ between
the normal to the Earth's surface (passing
through the detector) and the direction of 
the neutrino beam incident on the detector
($\theta=0^o$ corresponds to a beam traversing
the diameter of the Earth).  It is obtained
from integrating the density $\rho(r)$ of 
Earth \cite{ref49} along the neutrino beam
path at a given $\theta$ and is given in 
Fig.\ 15 of \cite{ref20} where $X(\theta)$
has been denoted by $z(\theta)$.  For
definiteness all above formulae have been
given for an incoming neutrino beam, but
similar expressions hold of course for 
antineutrinos.

Before evaluating total event rates, we first
compare the effects of attenuation and 
regeneration in (2.10) for the initial 
differential $\nu +\bar{\nu}$ fluxes in 
Fig.\ 1 when they reach the Earth's surface 
($\Phi_{\nu,\bar{\nu}}(E_{\nu})=\Phi
(E_{\nu})/4$), with the pure attenuation 
(absorption) where regeneration is omitted 
($\Psi_{\nu,\bar{\nu}} \equiv 1$ in (2.10)).  
Some typical 
representative results for the differential
shadow factor in (2.10) are shown in Fig.\ 3
where, for comparison, the pure attenuation
(absorption) is shown by the dashed curves.
The results for $\nu$ and $\bar{\nu}$ fluxes
are shown separately in Fig.\ 3(a) and
Fig.\ 3(b), respectively, which are combined
in Fig.\ 3(c) for the total fluxes in Fig.\ 1.
For the rather flat (initial) cosmic fluxes
in Fig.\ 1, the regeneration effect is 
significant and increases the fluxes at the
detector by as much as 30\% for $\theta = 0^o$
and about 10\% for $\theta =40^o$ as compared
to the purely absorbed (attenuated) fluxes.
In fact it can even enhance the initial flux
at lower energies ($E_{\nu}$ 
\raisebox{-0.1 cm}{$\stackrel{<}{\sim}$} $10^4$
GeV) at $\theta$ \raisebox{-0.1 cm}{$\stackrel
{<}{\sim}$} $40^o$ as shown for the AGN--SS 
flux in Fig.\ 3.  A similar result was 
originally found in \cite{ref44} investigated
in more detail in \cite{ref35} for (even
somewhat flatter) actual neutrino fluxes.
Below $10^6$ GeV the shadow factors for 
antineutrinos in Fig.\ 3(b) are obviously 
larger than the ones for neutrinos in 
Fig.\ 3(a), since $\sigma^{\bar{\nu}N}<
\sigma^{\nu N}$, except for the regenerated
AGN--SS flux at $\theta =0^o$ below $10^4$
GeV where, in addition, AGN--SS strongly
decreases for decreasing $E_{\nu}$ as can be
seen in Fig.\ 1.  At larger energies ($E_{\nu}$
\raisebox{-0.1 cm}{$\stackrel{>}{\sim}$} 
$10^6$ GeV), where the CC and NC cross sections 
for neutrinos and antineutrinos practically 
coincide, the shadow factors in Figs.\ 3(a) 
and 3(b) become indistinguishable.
Clearly, once the energy is sufficiently  
high the attenuation factor $\exp (-N_A
\sigma^{\stackrel{(-)}{\nu}\!N}_{\rm tot}X)$
leads to total shadowing as seen in Fig.\ 3 
due to the increase of (anti)neutrino cross
sections with energy.  The smaller the nadir
angle the lower the energy of complete
attenuation.

For experimental purposes the relevant and
interesting quantity is the total 
$\nu+\bar{\nu}$ flux
$\Phi_{\nu +\bar{\nu}}(E_{\nu},X)$ 
reaching the detector, after its passage
through Earth, at different nadir angles
$\theta$.  The four plots in Figs.\ 4(a)
and 4(b) show first the initial 
$\nu +\bar{\nu}$ fluxes reaching the Earth's
surface, $\Phi_{\nu +\bar{\nu}}(E_{\nu},X
=0)\equiv \Phi_{\nu +\bar{\nu}}(E_{\nu})=
\Phi(E_{\nu})/2$ with the original cosmic fluxes
$\Phi$ being given in Fig.\ 1, and then the
fluxes at the detector 
$\Phi_{\nu +\bar{\nu}}(E_{\nu},X)$ for 
the three nadir angles $\theta =80^o,\, 
40^o$ and $0^o$.  (Recall that $0^o$ is 
the passage of (anti)neutrinos through the
center of the Earth.)  Due to the large
$\stackrel{(-)}{\nu}\!\!\!N$ cross sections
the attenuation effects (absorption plus
regeneration) reduce dramatically the 
fluxes of ultrahigh energy (anti)neutrinos,
particularly at small nadir angles in 
Fig.\ 4.  For convenience of reference the 
atmospheric (ATM) neutrino flux is shown in
both Figs.\ 4(a) and 4(b).  The AGN fluxes in
Fig.\ 4(a) stand out above this background ATM
neutrino spectrum for neutrino energies above
about $10^5$ GeV, whereas the flatter TD and
$Z$--burst fluxes in Fig.\ 4(b) are overwhelmed
by the ATM background below $10^6$ GeV, in
particular for smaller nadir angles. Nevertheless there remains a 
window for the observation of cosmic
neutrinos (with $E_{\nu}$ 
\raisebox{-0.1cm}{$\stackrel{<}{\sim}$}
$10^8$ GeV) by underground detection of 
the energetic $\mu^{\pm}$ when calculating
total event rates.

The total upward $\mu^+ +\mu^-$ event rates
are calculated according to (2.1) for the
(forthcoming) AMANDA--II \cite{ref2,ref50},
IceCube \cite{ref51} and ANTARES 
\cite{ref39} detectors with their respective 
energy dependent areas $A(E_{\mu})$ shown in 
Fig.\ 5. The required total $\nu_{\mu}$ and
$\bar{\nu}_{\mu}$ flux in (2.1) reaching
the Earth's surface is given by 
$\Phi_{{\nu}_{\mu}+\bar{\nu}_{\mu}}\equiv
\Phi_{{\nu}_{\mu}}+\Phi_{\bar{\nu}_{\mu}}
=\Phi/2$, with the original cosmic neutrino
fluxes $\Phi$ given in Fig.\ 1, since the
recent discovery of near--maximal $\nu_e-
\nu_{\mu}$ and $\nu_{\mu}-\nu_{\tau}$ mixing 
\cite{ref52} implies that the originally
produced cosmic neutrino fluxes, having a
$\nu_e:\nu_{\mu}:\nu_{\tau}$ ratio of $1:2:0$
at some astrophysical source, inevitably 
oscillate \cite{ref53,ref54} to a ratio of
$1:1:1$.  The resulting annual total 
nadir--angle--integrated rates are given in
Table 1 where the numbers in parentheses are
the event rates with no regeneration ($\Psi
\equiv 1$ in (2.10)), i.e.\ the attenuation
of the Earth's penetrating neutrinos is just
due to naive absorption.  Regeneration effects
increase the event rates on the average by
about 20\% as compared to the ones where the
attenuation of neutrinos is just caused by
absorption \cite{ref20,ref21,ref23}.  The
contribution to the event rates in Table 1
from energies above $10^8$ GeV becomes, however,
negligible and unmeasurably small due to the 
reduction of the initial neutrino fluxes by
attenuation with or without regeneration.  The
highest signal rates arise in the AGN models
which might be testable for neutrino flux
energies as large as $10^7$ -- $10^8$ GeV, i.e.\
$E_{\mu}^{\rm min}=10^7$ GeV. Beyond neutrino
energies of $10^8$ GeV, however, present models 
of cosmic neutrino fluxes are not testable by
upward--going $\mu^++\mu^-$ events. Notice that 
the atmospheric neutrino background becomes
marginal for neutrino energies above 
$10^{5}$ GeV \cite{ref20,ref21,ref35}, i.e.\ 
$E_{\mu}^{\rm min}=10^5$ GeV in Table 1, 
or in other words the ATM rate comes entirely
from $E_{\nu}<10^6$ GeV.
Furthermore the energy dependence of the
upward--going muon rate in Table 1 will be an
important discriminant for separating 
atmospheric and extraterrestrial sources.
The nadir angular dependence of the annual
upward $\mu^+ +\mu^-$ event rates shown in
Fig.\ 6 for two representative cosmic neutrino
flux models illustrates in more detail the
enhancement of event rates caused by 
regeneration.  Apart from the absolute
normalizations, these angular distributions
are not too sensitive to the chosen value of
$E_{\mu}^{\rm min}$ and clearly favor shallow
nadir angles, i.e.\ large $\theta$, where  
the largest amount of events reside.  The
dashed histograms in Fig.\ 6 refer to events
where the neutrino attenuation is caused just
by absorption ($\Psi\equiv 1$ in (2.12)) with
regeneration effects disregarded which give
rise to the total event rates in parentheses
in Table 1.  Due to maximal 
$\nu_{\mu}-\nu_{\tau}$ mixing, the $\nu_{\tau}+
\bar{\nu}_{\tau}$ flux arriving at the Earth's
surface may enhance these upward $\mu^+ +\mu^-$
rates due to their interaction in Earth via
$\nu_{\tau}N\to\tau X\to\mu X'$ 
\cite{ref55,ref56,ref57}. Here we disregard such 
additional contributions since Earth--skimming
tau--neutrinos \cite{ref42,ref43,ref25} will 
dominate over these interactions and allow to 
test cosmic neutrino flux models at highest 
energies to which we shall turn later.

The upward event rates studied thus far depend
obviously rather little \cite{ref20} on the
specific choice of parton distributions since
the combination $P_{\mu}S$ in (2.3) is rather
insensitive with respect to different choices:
for increasing energies the effect of larger
cross sections is to increase the probability
$P_{\mu}$ that a neutrino produces an observable
muon, but also to increase the attenuation of
neutrinos via $S$ on their way to the detector.
This is of course in contrast to the downward
event rates for $\stackrel{(-)}{\nu}$'s 
that enter the detector from above or 
quasi--horizontally, i.e.\ where \cite{ref20}
$S\simeq 1$.  For very high muon--energy
thresholds it will thus be necessary to observe
downward--going muons produced by interactions
within the instrumented volume.  Our expected
annual
downward $\mu^+ +\mu^-$ event rates are shown
in Table 2 which are of course consistent with
the ones observed in \cite{ref21} for 
appropriate fluxes and parameters used there.
These results are encouraging and allow to test
cosmic neutrino fluxes at higher neutrino
energies (by about a factor of 10 higher than
for the upward rates in Table 1), provided that
downward--going contained events can be observed 
efficiently. 
For comparison we show again in Table 2 the 
background atmospheric ATM rates which, for the 
sizeable and measurable cosmic rates, play no 
role for $E_{\nu}$ 
\raisebox{-0.1cm}{$\stackrel{>}{\sim}$} $10^6$ GeV.
Despite the fact that underground muon detectors
(will) have a poor energy resolution, a measurement
of the energy dependence of the downward--going
muon rate will be again an important discriminant
for separating atmospheric and extraterrestrial
sources below $10^6$ GeV.  (Notice that the 
downward muon rates in Tables 2 and 3 remain
essentially unchanged if the detector depth is 
taken into account.  Since underground detectors
are deployed at a depth of 2 to 3 km, the limited
amount of matter above the detector does not reduce
the shadow factor $S(E_{\nu})\simeq 1$ in (2.3) 
which holds for all relevant energies in Tables
2 and 3.  Only at energies as high as 
$E_{\nu}=10^{10}$
GeV it decreases to $S=0.99\, .$)

Here the question arises whether some other
reasonably founded extrapolation to the 
ultrasmall--$x$ region could sizeably enhance
our benchmark rates (e.g.\ in \makebox{Table 2)}
 calculated
from cross sections as derived from our nominal
QCD--dynamical GRV98 parton distributions (or
equivalently from the appropriately extrapolated
CTEQ3-DIS parametrization). In particular for 
  `new physics' searches 
\cite{ref1,ref3,ref4,ref5,ref13,ref58,ref59,ref60}
it will be important to know to what extent the
conventional standard model would allow for 
(sizeably) larger rates.  For this purpose we
employ a Regge model inspired small--$x$
description of the DIS ep structure function
$F_2^{ep}(x,Q^2)$ recently suggested by 
Donnachie and Landshoff (DL) \cite{ref61}.
According to DL, $F_2^{ep}$ may be written as
a sum of three factorized terms 
$f_i(Q^2)x^{-\varepsilon_i}$ to be fitted to
DIS HERA data as well as to total photoproduction
$(Q^2=0)$ cross sections.  This result may be
used as a (possibly extreme) guideline for an
extrapolation into the ultrasmall--$x$ region,
dominated by the `hard pomeron' component 
$\varepsilon_0\simeq 0.4$, and interpolating
\cite{ref62} it smoothly  to measured 
(anti)neutrino structure functions 
$F_{2,3}^{\stackrel{(-)}{\nu}N}(x,Q^2)$ by
utilizing the CTEQ5 parametrization gives rise
to a quicker power--like increase of 
$\stackrel{(-)}{\nu}\!\!\!N$ cross sections at 
extremely high energies, $E_{\nu}$
\raisebox{-0.1cm}{$\stackrel{>}{\sim}$} $10^8$
GeV, due to the power--like small--$x$ behavior
implied by the DL--fit.  This approach will be
called \cite{ref62} DL+CTEQ5 hereafter.  The
resulting $\sigma_{\rm tot}^{\nu N}(E_{\nu})$ is
shown in Fig.\ 7 and compared with our nominal
(radiative GRV98 or CTEQ3--DIS) cross sections
as well as with the somewhat smaller one
corresponding to CTEQ4--DIS. The Regge  `hard
pomeron' pole small--$x$ extrapolated DL+CTEQ5
cross section is, at highest energies, a factor
of about 2 larger than our nominal cross 
section predicted by purely radiative QCD--RG
evolutions which may be considered as a 
reasonable upper bound implied by a 
`conventional' standard model approach.  
Similar remarks hold for the resulting 
downward $\mu^+ +\mu^-$ event rates in Table 3
which should be compared with the nominal ones
in Table 2: typically the Regge--inspired rates
are enhanced by about 20\%.  It appears that
yet much larger rates and cross sections than
those in Table 3 and Fig.\ 7 are unlikely to
be accommodated by currently conceivable 
standard model approaches and, if confirmed
by future measurements, might require  `new'
physics ideas for their explanation.

\setcounter{equation}{0}
\section{Event Rates for Surface Telescopes} 
For exploring the entire spectrum of cosmic
neutrino fluxes up to highest neutrino energies
of about $10^{12}$ GeV in Fig.\ 1, large--area
ground arrays and/or surface air fluorescence
telescopes (like the Pierre Auger Observatory
\cite{ref28,ref63} and the Telescope Array
\cite{ref26,ref64}), where the interaction
medium is not the Earth but the atmosphere, 
will become instrumental since upward--going
neutrinos are essentially blocked by Earth 
for energies above $10^8$ GeV.  Here,
near--horizontal incoming neutrinos will produce
electromagnetic and/or hadronic extensive air
showers which will be observed by \v{C}erenkov
radiation and/or fluorescence detectors.  For
a $(\nu+\bar{\nu})$--flux $\Phi_{\nu+\bar{\nu}}$
reaching Earth, the event rate per second 
for deeply penetrating horizontal showers 
is given by \cite{ref28,ref26}
\begin{equation}
\rm{rate} \left[E_{\rm sh}>E_{\rm th}\right]
 = N_A\, \rho_{\rm air} \int_{E_{\rm th}}
    dE_{\rm sh} \int_0^1 dy\, 
\frac{d\Phi_{\nu+\bar{\nu}}(E_{\nu})}{dE_{\nu}}\,\,
\frac{d\sigma^{(\nu+\bar{\nu})N}(E_{\nu},y)}{dy}
       \, {\cal{A}}(E_{\rm sh})
\end{equation}
where $\rho_{\rm air}\simeq 10^{-3}$ g/cm$^3$
and $\Phi_{\nu_{\ell}+\bar{\nu}_{\ell}}=\Phi/2$
for $\ell = e,\,\mu,\,\tau$ and with the
original $\nu_{\mu}+\bar{\nu}_{\mu}$ flux 
$\Phi$ at its production site given in Fig.\ 1
and the electron neutrino flux has been 
approximated by naive channel counting in pion
production and decay ($\nu_e/\nu_{\mu}=1/2$)
together with maximal mixing.  The relation
between the shower energy $E_{\rm sh}$ and the
primary neutrino energy $E_{\nu}$ depends on
the neutrino interaction being considered 
and ${\cal{A}}(E_{\rm sh})$ is the (geometric)
detector acceptance.  If a detector just
measures the shower energy (i.e.\ cannot
distinguish between hadronic and 
electromagnetic showers) we have $E_{\rm sh}
=E_{\nu}$ for $(\nu_e+\bar{\nu}_e)N$ CC
interactions, whereas $E_{\rm sh}=yE_{\nu}$
for $(\nu_{\mu}+\bar{\nu}_{\mu})N$ CC 
interactions where electromagnetic showers
are negligible.  This latter relation 
$E_{\rm sh}=yE_{\nu}$ holds also for all NC
reactions which always produce hadronic
showers. If a detector can distinguish between
hadronic and electromagnetic showers this 
latter relation remains the same for all cases
discussed except for $(\nu_e+\bar{\nu}_e)N$
CC interactions where the neutrino energy has
now to be shared between the hadronic and 
electromagnetic shower energy, i.e. 
$E_{\rm{sh},\,h}=yE_{\nu}$ and 
$E_{\rm sh,\,elm}=(1-y)E_{\nu}$.
In Fig.\ 8 we display the relevant acceptances
for the Auger detector \cite{ref28} used for
our calculations.  We shall not employ the 
Auger acceptance as calculated by Billoir
\cite{ref65} which has only been presented up
to $E_{\rm sh} = 10^{10}$ GeV since we refrain
from arbitrarily extrapolating \cite{ref21}
it to $10^{12}$ GeV. The expected
annual event rates are displayed in Table 4,
employing the geometrical acceptances of 
Fig.\ 8.  Utilizing instead the Monte Carlo
simulated acceptance in Fig.\ 8 the resulting
total (CC + NC) Auger rates would be about a
factor of 3 smaller than the ones shown in
Table 4.  The largest event rates arise from
$(\nu_e+\bar{\nu}_e)N$ CC interactions, for 
which $E_{\rm sh}\simeq E_{\nu}$.  Therefore
a measurement of the shower--energy dependence
will shed some light on the theoretical cosmic
neutrino flux models in Fig.\ 1 at ultrahigh
energies $E_{\nu}>10^8$ GeV despite their
decrease with increasing $E_{\nu}$ which, in
some cases (TD, $Z$--burst), is overcompensated
by the increasing acceptances in Fig.\ 8.
For example, quasi--horizontal air showers
with $E_{sh}>10^8$ GeV will test the AGN--M95
flux dominantly at $E_{\nu}\simeq 10^8$ --
$10^{10}$ GeV, the TD--SLBY flux at about
$10^9$ -- $10^{11}$ GeV and the $Z$--burst
initiated flux at $10^{10}$ -- $10^{12}$ GeV
according to the rates for the individual
$E_{\rm sh}$--bins in Table 4.  Unfortunately
the total amount of air--shower events is 
modest and in one year a few events may be
observed in the Auger detector assuming modern
estimates of AGN--M95, TD-SLBY and $Z$--burst
neutrino fluxes.  Most other cosmic neutrino
flux models, in particular the recently 
updated and corrected GRB--WB flux \cite{ref11},
yield fractions of one event per year and 
remain undetectable by horizontal air showers.
This situation remains practically unchanged
even for the optimal Regge--inspired structure
functions resulting in the largest $\nu N$
cross section in Fig.\ 7 (event rates shown
in parentheses in Table 4) which increase our
nominal rates by less than 50\%. (Our results
agree of course with the total air shower
rates estimated in \cite{ref21,ref28} when
using the fluxes, structure functions and 
acceptances employed there.)  Our total Auger
results in Table 4 are compared, where
available, with the expected total rates for
the TA detector \cite{ref26} which in some
cases may be about a factor of 2 larger than
for Auger. 

Furthermore, contributions from 
$\tau$--neutrinos, with similar acceptances
above $10^8$ GeV as for $\mu$--neutrinos
\cite{ref26}, may enhance the total air shower
rates by as much as 20 to 30\% \cite{ref26}.
We disregard such contributions here, partly
because the appropriate detector acceptances
for $\nu_{\tau}$ induced air showers do not
exist in the literature, but mainly because
Earth--skimming tau--neutrinos 
\cite{ref42,ref43,ref25} will dominate over
atmospheric interactions.  

Tau--neutrinos when skimming the Earth
\cite{ref42,ref43,ref25} will be far more effective
in producing, besides electromagnetic showers,
specific double shower (`double bang') events.
When ultrahigh energy cosmic neutrinos 
($E_\nu$ \raisebox{-0.1cm}{$\stackrel{>}{\sim}$}
$10^8$ GeV) penetrate and skim the surface
of Earth they convert to charged leptons
which may leave the Earth's surface essentially
horizontally.  For this to happen the neutrinos
have to enter Earth near--horizontally at some
critical nadir angle $\theta(E_{\nu})$, 
typically \cite{ref25,ref26} $\theta(E_{\nu})$
\raisebox{-0.1cm}{$\stackrel{>}{\sim}$} $85^o$,
where they travel along chords with length of
the order of their charged current interaction
length $\lambda_{\rm int}^{CC}(E_{\nu})$. For
 larger
nadir angles they rarely interact to produce
charged leptons, whereas for smaller ones the 
neutrinos are shadowed by Earth.  Electrons
lose their energy too quickly in the Earth for
being detected by ground--level surface 
telescopes.  Furthermore, since $\beta_{\tau}
\simeq 0.8\times 10^{-6}$ cm$^2$/g $\ll
\beta_{\mu}\simeq 6.0 \times 10^{-6}$ cm$^2$/g
in the energy--loss relation (2.6), muons and
taus can travel 1.5 km and 11 km, respectively,
before losing a decade in energy 
\cite{ref25,ref48}.  Thus, tau--neutrinos 
contribute dominantly to Earth--skimming events
observed above the Earth's surface which 
consist mainly of electromagnetic showers and
of  `double bang' showers 
\cite{ref25,ref26,ref42,ref43}.  The detected
tau leptons determine the relevant total 
event rate which is given by \cite{ref25}
\begin{equation}
N_{\tau}
 = 2\int dE_{\nu}\, dE_{\tau}\,d\varphi\, d\cos\theta\,
  K(E_{\nu},\theta;\ E_{\tau})\, 
   \frac{\cos\theta}{2\pi}\,\,
    \frac{d\Phi_{\nu_{\tau}}(E_{\nu})}{dE_{\nu}}
     \, A_{\rm eff}^{\Omega}(E_{\tau})\rm{TD}
\end{equation}
for an isotropic neutrino flux $\Phi_{\nu_{\tau}}
=\Phi_{\bar{\nu}_{\tau}}=\Phi/4$, according to
the definition below Eq.\ (3.1), arriving with
nadir angles $\theta<\pi/2$ and the additional
$\cos\theta$ derives from projecting its 
trajectory (which nearly coincides with the 
$\tau$--trajectory for UHE $\nu_{\tau}$'s) onto
the surface area \cite{ref66}.  The probability
(per $E_{\tau}$) that a $\nu_{\tau}$ entering
Earth with energy $E_{\nu}$ and nadir angle
$\theta$ produces a $\tau$ that exits Earth
with energy $E_{\tau}$ is
\begin{eqnarray}
K(E_{\nu},\theta;\,E_{\tau})\,  & \simeq  &
 N_A\sigma_{CC}^{\nu N}(E_{\nu}) \,
  \frac{1}{E_{\tau}\beta_{\tau}}\, \exp 
   \left[ -N_A\sigma_{CC}^{\nu N}(E_{\nu})
    \int_0^{2R_{\oplus}\cos\theta}dz'
     \rho\left(r(\theta,z')\right)\right]
\nonumber\\
  & & \quad\quad\quad\quad\quad\quad
       \times\quad \exp
        \left[\frac{m_{\tau}}
         {c\tau_{\tau}\beta_{\tau}\rho_{s}}
          \left(\frac{1}{E_{\nu}}
           -\frac{1}{E_{\tau}}\right)\right]
\end{eqnarray}
with $\rho(r)$ being the Earth density at distance 
$r$ from its center (taken from Fig.\ 14 of
\cite{ref20}) with $r^2(\theta,z)=R_{\oplus}^2
+z^2-2R_{\oplus}z\cos\theta$, $R_{\oplus}\simeq
6371$ km, where $z$ is the distance the incoming
$\nu_{\tau}$ travels through Earth before
converting to a $\tau$ near the Earth's surface
where the density is $\rho_s\simeq 2.65$ g/cm$^3$.
Furthermore, $c\tau_{\tau}=87.11 \,\mu$m and
$m_{\tau}=1777.03$ MeV.  To account for the
requirement of clear moonless nights for 
fluorescence detection, a duty cycle of 10\%
will be assumed in (3.2) as usual, $D=0.1$, 
and the time $T$ an experiment runs is taken,
for convenience, to be 1 year although, once
approved, the experiment will run for several
years.  For reasons discussed above, the 
integration over the nadir angle will be 
restricted to \cite{ref25,ref26} $90^o-\theta
\leq 5^o$ although the results do not depend
too strongly on this specific choice.  The
factor of 2 in front of the integral in (3.2)
accounts for the contribution of incoming
antineutrinos since for $E_{\nu}$
\raisebox{-0.1cm}{$\stackrel{>}{\sim}$} 
$10^{8}$ GeV the CC $\nu_{\tau}$ and
$\bar{\nu}_{\tau}$ cross sections are virtually
identical.  The effective (geometrical) aperture
$A_{\rm eff}^{\Omega}(E_{\tau})$ in (3.2) as
estimated \cite{ref25} for the TA detector
(one station) is shown in Fig.\ 9 where for
comparison the one for the far smaller HiRes
detector is shown as well.  The sensitivity of
the fluorescence detectors of the Auger
Observatory is expected \cite{ref25} to lie
somewhere between that of HiRes and TA in 
Fig.\ 9, but such a quantity has unfortunately
not yet been separately published.

The nominal event rate (based on our GRV98 or
CTEQ3--DIS parton densities) for the six 
neutrino sources given in Fig.\ 1, binned by 
the $\tau$--energy, are given in Table 5
for the TA detector assuming 10 observational
stations \cite{ref26}. (We refrain from 
recalculating the rates for the Fly's Eye and
HiRes detectors which have been shown to be
marginal \cite{ref25}.)  The expected rates 
for the \underline{largest} Regge--inspired
$\nu N$ cross section (cf.\ Fig.\ 7) are shown
in parentheses which are in most cases 
\underline{smaller} than the nominal ones 
(based on GRV98 or CTEQ3-DIS in Fig.\ 7) since
the first (attenuation) exponential in (3.3)
becomes dominant for $E_{\nu}$
\raisebox{-0.1cm}{$\stackrel{>}{\sim}$} $10^9$
GeV even for the rather small column depth 
involved.  This implies that if event rates are
calculated utilizing even smaller UHE $\nu N$
cross sections than the ones in Fig.\ 7 (due 
to flatter and thus less steep 
extrapolations to the ultrasmall Bjorken--$x$
region $x\sim M_W^2/s$ of conventionally fitted
parton densities), one easily 
\underline{over}estimates Earth--skimming rates
at highest energies.  For comparison the total
TA event rates as estimated in \cite{ref26} are,
where available, shown as well which are 
somewhat smaller than our results and the ones
obtained in \cite{ref25}.  This is presumably
due to the fact that the realistic TA aperture is
expected to be somewhat smaller \cite{ref67} 
than the one
calculated in \cite{ref25} and shown in 
Fig.\ 9 because of effects
such as the detector response, light propagation
in air and night sky background.  The total 
rates for the Auger detector are also shown in
Table 5 which we have calculated for our 
neutrino fluxes in Fig.\ 1 using the effective
Auger aperture \cite{ref43} where, besides the
dominant electromagnetic showers, multi--bang
events have been also taken into account for
the Monte Carlo simulations (for definiteness
we used the average aperture in Fig.\ 9 of
\cite{ref43} as obtained from the `BS+PP+DIS low'
continuous energy loss model).  In any case,
in contrast to the quasi--horizontal air showers
in Table 4, the shower event rates produced by
Earth--skimming $\nu_{\tau}$'s are sufficiently
large for testing and exploring the large
energy tail (\raisebox{-0.1cm}{$\stackrel{>}
{\sim}$} $10^8$ GeV) up to $10^{11}$ GeV of
most cosmic neutrino sources in Fig.\ 1, except
for the recently updated and corrected GRB--WB
flux \cite{ref11} for which the rates remain
prohibitively small.

In order to learn to what extent the shower
rates measured in certain bins of $E_{\tau}$ as
shown in Table 5 can explore and test the 
energy profile of the theoretical cosmic 
neutrino fluxes $\Phi(E_{\nu})$ in Fig.\ 1, we 
display in Table 6 the individual 
$E_{\nu}$--binned contributions to our nominal
TA--rates for the $E_{\tau}$--bins given in 
Table 5 for three representative neutrino fluxes.
Apart from the less steep falling $Z$--burst
flux in Fig.\ 1 which contributes over a wider
$E_{\nu}$--range for a given $E_{\tau}$--bin,
measured showers with $E_{\tau}=10^8-10^9$ GeV
will test cosmic neutrino fluxes mainly within
$E_{\nu}=10^8-10^{10}$ GeV, whereas increasing
the shower energy to $E_{\tau}=10^9-10^{10}$
GeV, for example, will delineate a similar
$E_{\nu}$--bin, $E_{\nu}=10^9-10^{10}$ GeV.
The same holds true for even higher values of
$E_{\tau}$.  Therefore from an experimental
point of view it will be important to measure
the detailed energy ($E_{\tau}$) dependence
predicted in Table 5 for exploring cosmic
neutrino sources at highest energies.

\setcounter{equation}{0}
 \section{Implications of Low Scale `Extra
Dimensions'}
Having calculated so far in the previous Section
the muon and shower
event rates caused by extragalactic neutrinos
according to conventional (nominal) and 
optimal (Regge inspired) standard model
expectations, we will now finally give for
comparison a few examples of distinctively 
different event rates which derive from  `new'
physics ideas.  Although highly speculative and
by far non--unique, let us choose string theories
with large  `extra dimensions' $\delta$ where
only gravity propagates in the $4+\delta$ 
dimensional bulk of spacetime (a recent brief
review can be found in \cite{ref68}). To obtain
the effective low--energy theory in $3+1$
dimensions, these extra dimensions are 
compactified to a common scale $R$ which may
be relatively large, corresponding to a small
scale $1/R$ of new physics, and is related to
standard Newton gravity (Planck scale) via
$G_N^{-1}\sim M_S^{\delta+2}R\,^{\delta}$
with $M_S\sim$ TeV being the low energy 
effective string scale 
\cite{ref69}.  (For example, $R$ is of the
order of 1 mm for $\delta=2$.)  In addition
to the usual Standard Model (SM) particles,
an infinite tower of massive Kaluza--Klein
(KK) excitations appears in the effective
four--dimensional theory, corresponding to 
the massless graviton in $4+\delta$ dimensions,
and $M_S$ plays the role of an ultraviolet
cutoff for the summation over all relevant
spin--2 KK states.  Such a scenario opens up
the interesting possibility of massive spin--2
$t$--channel exchange which results in a more
rapid growth with energy as compared to SM
cross sections derived from massive vector
boson exchange considered thus far.  Being
flavor--neutral this  `graviton' KK exchange
($G$) contributes only via NC processes 
to the total SM $\nu N$ cross section.
The cross sections for the partonic subprocesses
$\nu\!\!\!\stackrel{(-)}{q}\to
\nu\!\!\!\stackrel{(-)}{q}$
and $\nu g\to\nu g$ as mediated by $G$ read
\cite{ref58,ref59,ref70,ref69}
\begin{eqnarray}
\frac{d\hat{\sigma}_G^{\nu\!\!\!\stackrel{(-)}{q}}}
  {d\hat{t}} & = & \frac{\pi}{32 M_S^8}\,\,
   \frac{1}{\hat{s}^2}
    \left[ 32\hat{s}^4+64\hat{s}^3\hat{t}+
     42\hat{s}^2\hat{t}^2+10\hat{s}\hat{t}^3+\hat{t}^4
      \right] F(\hat{s},\hat{t}\,)\\
\frac{d\hat{\sigma}_G^{\nu g}}{d\hat{t}}
     & = & \frac{\pi}{2M_S^8}\,\,
      \frac{1}{\hat{s}^2}
       \left[ 2\hat{s}^4+4\hat{s}^3\hat{t}+
        3\hat{s}^2\hat{t}^2+\hat{s}\hat{t}^3
         \right] F(\hat{s},\hat{t}\,)
\end{eqnarray}
for $\delta =2$ extra dimensions and the terms in
square brackets are symmetric under 
$\hat{s}\leftrightarrow\hat{u}$.  The function 
$F(\hat{s},\hat{t}\,)$ refers to some 
unitarization (interpolation) procedure
\cite{ref58,ref59,ref71} in order to extrapolate
the perturbative predictions at $\hat{s}$
\raisebox{-0.1cm}{$\stackrel{<}{\sim}$} $M_S^2$
in (4.1) and (4.2) to $\hat{s}\gg M_S^2$.
We shall use two representative choices
\cite{ref58},
\begin{eqnarray}
F(\hat{s},\hat{t}\,) & = & 
 \frac{M_S^6}{(M_S^2+\hat{s})^2(M_S^2-\hat{t})}
  \,\left[1+10\ln (1+\hat{s}/M_S^2)\right]\\
F(\hat{s},\hat{t}\, ) & = & 
  \frac{M_S^4}{(M_S^2+\hat{s})(M_S^2-\hat{t})}
\end{eqnarray}
as well as choosing $M_S=1$ TeV for our subsequent
quantitative analysis.  The contribution to the
total SM $\nu N$ cross section due to  `graviton'
KK exchange is thus given by
\begin{equation}
\frac{d\sigma_G^{\nu N}}{dx\,dy} =
  s\sum_{f=q,\,\bar{q},\,g} xf(x,Q^2)
   \frac{d\hat{\sigma}_G^{\nu f}}{d\hat{t}}
\end{equation}
with $\hat{s}=xs$, $\hat{t}=t\equiv -Q^2=-xys$
and only the dominant light quarks $q=u,\,d,\,s$
are taken into account.  The same expression
holds for $\bar{\nu} N$ scattering.  In Fig.\ 10
our nominal SM total (CC + NC) $\nu N$ cross
section is compared with the additional KK
contribution where the dominant gluon initiated
component in (4.5) is, at highest energies, 
about a factor of 2 to 4 larger than the sea
contributions of (anti)quarks depending on 
whether the interpolation (4.4) or (4.3), 
respectively, is used.  
(It should be noticed that these cross sections
are in agreement with the upper bound derived
in \cite{ref72}.  A more recent updated estimate
\cite{ref73}, however, being based on the full
exposure of the AGASA and Fly's Eye experiments
as well as on a larger cosmogenic neutrino flux
\cite{ref74} together with a larger nucleon
source cutoff energy, would rule out the largest
strongly rising cross section (dashed curve)
in Fig.\ 10 at highest neutrino energies.)
These total SM + KK
$\nu N$ cross sections can rise to hadronic
mb--scale values \cite{ref58,ref59} at highest
neutrino energies and the resulting annual 
total event rates for the relevant 
quasi--horizontal air showers, shown in Table 7,
are enormous when compared with the small SM
rates in Table 4.  Even for highest neutrino
energies (i.e.\ highest threshold energies in
(3.1)) the number of events remains large,
except for the steeply falling AGN-SS and 
GRB--WB fluxes in Fig.\ 1.  It should be easy
for the future Pierre Auger Observatory to 
observe such dramatically large hadronic shower
event rates as implied by low--scale large
 `extra dimension' scenarios.  
(Given the uncertainties in the cosmic neutrino
flux predictions, the eventual detection of 
such event rates will be subject to interpretation,
because they are proportional to the product of
the flux and the cross section.  Ways to
 discriminate between large neutrino cross
sections and large neutrino fluxes have been
suggested recently by looking at the zenith
angle distribution of the neutrino induced
shower events, i.e.\ by comparing the energy
spectrum of upgoing and nearly horizontal
to downgoing shower events \cite{ref66,ref75}.
Such a measurement may at least in principle
allow to determine $\sigma^{(\nu+\bar{\nu})N}$ 
independently of the neutrino flux.)
The viability
of such  `new' physics scenarios may be even
tested by the smaller ground level AGASA
\cite{ref76} and Fly's Eye (HiRes) 
\cite{ref77} observatories: using their 
estimated shower acceptances 
\cite{ref78,ref77}, the expected rates for 
AGASA are about a factor of 25 smaller than the
ones in Table 7, and a similar reduction holds
for the Fly's Eye (HiRes) observatory.

\section{Summary}
In order to explore and test the sources of
cosmic UHE neutrinos, the calculation of the
relevant $\stackrel{(-)}{\nu}\!\!\!N$ cross sections
requires the knowledge of parton distributions
$f(x,Q^2)$, with $f=q,\,\bar{q},\,g$, at
ultrasmall Bjorken $x\sim M_W^2/s$, below the
range constrained by present experiments, and
at scales $Q^2=M_W^2$.  Highest neutrino
energies $E_{\nu}=s/2M_N\simeq 10^{12}$ GeV
require extrapolations down to $x\simeq 10^{-9}$.
Within standard QCD RG--evolutions, such 
extensive small--$x$ extrapolations can be 
uniquely and reliably predicted by the dynamical
(radiative) parton model \cite{ref32} which
proved to provide reliable high energy small--$x$
predictions in the past.  (Interestingly, 
consistent BFKL model resummations in leading
order QCD yield remarkably similar results
\cite{ref35}.) All resulting relevant
$\stackrel{(-)}{\nu}\!\!\!N$ cross sections 
turn out to have a typical uncertainty of about
$\pm$ 20\% at highest neutrino energies of
$10^{12}$ GeV \cite{ref36}.  We have adopted
this  `nominal' approach for all our calculations
which are based on the canonical radiative 
GRV98 parton distributions \cite{ref32}. 
Similar results are accidentally obtained 
from the conventionally fitted CTEQ3--DIS 
parametrizations \cite{ref37} at $x$
\raisebox{-0.1cm}{$\stackrel{>}{\sim}$} $10^{-5}$
with their assumed fixed--power extrapolation
to $x<10^{-5}$, whereas the CTEQ4--DIS
parametrizations \cite{ref38} somewhat
underestimate the neutrino--nucleon cross
sections by about 20\% at $E_{\nu}\simeq 
10^{10}- 10^{12}$ GeV.  In calculating event 
rates caused by cosmic UHE neutrinos we have 
also employed recent Regge--model inspired 
small--$x$ structure functions 
\cite{ref61,ref62} in order to learn about
possible maximal rates which may be 
accommodated by  `conventional' standard
model approaches and which do not necessarily
require  `new' physics interpretations.

For the representative set of modern cosmic
neutrino fluxes in Fig.\ 1 we first analyzed
$\mu^++\mu^-$ event rates caused by 
upward--going UHE neutrinos for modern (future)
underground detectors (such as ANTARES,
AMANDA II and IceCube), with special emphasis
on NC regeneration effects which populate the
lower energy part of the attenuated flux 
spectra.
Regeneration increases the non--regenerated
event rates \cite{ref20,ref21,ref23} on the
average by as much as 20\%. In particular 
AGN models, yielding the highest signal rates,
might be testable in this way for neutrino
energies as large as $10^7 - 10^8$ GeV, i.e.\
well above $10^5$ GeV where the atmospheric
neutrino background becomes marginal.  In 
addition we demonstrated in detail in Table 1
to what extent high--$E_{\nu}$ bins of the 
original fluxes in Fig.\ 1 remain experimentally
accessible after their depletion and 
regeneration due to NC interactions.  
Measurements of these energy--dependencies 
will be crucial for exploring the nature of
various cosmic neutrino sources.

In contrast to these upward $\mu^++\mu^-$
rates, the downward--going muon rates depend
strongly on the specific choice of parton
distributions and would allow to test cosmic
neutrino fluxes at higher $E_{\nu}$ (about a
factor of 10 higher than for the upward rates,
cf.\ Table 2), provided that downward--going
muons, produced by interactions within the
instrumented volume, can be observed
efficiently.  The largest Regge--inspired
cross section(s) in Fig.\ 7 give rise to 
downward $\mu^++\mu^-$ event rates (Table 3)
which are about $20 - 30$\% larger than our
nominal ones in Table 2, and may be considered
as reasonable upper bounds of present standard
model approaches.

To explore the entire spectrum of cosmic
neutrino fluxes up to highest neutrino energies
of about $10^{12}$ GeV, one has to resort to
quasi--horizontal air showers and 
Earth--skimming $\nu_{\tau}$'s to be detected
by large area ground arrays and surface air
fluorescence telescopes (such as the Pierre
Auger Observatory and the Telescope Array),
since upward--going neutrinos are essentially
blocked by Earth for energies above $10^8$
GeV.  The detailed shower energy $E_{\rm sh}$
dependence, which practically coincides with
the one of the incoming cosmic neutrinos, of
the annual event rates for the Auger detector
has been presented in Table 4 as well as the
results derived from the largest 
Regge inspired cross sections which can be
at most 50\% larger than our nominal rates.
The most efficient way, however, to test
cosmic neutrino fluxes at highest energies,
$10^8$ \raisebox{-0.1cm}{$\stackrel{<}{\sim}$}
$E_{\nu}$ \raisebox{-0.1cm}{$\stackrel{<}
{\sim}$} $10^{11}$ GeV, will be Earth--skimming
tau--neutrinos \cite{ref42,ref43,ref25} which, 
when converting to $\tau$--leptons that escape 
Earth, produce dominantly near--horizontal
electromagnetic showers as well as specific
 `double--bang' events.  Our nominal rates
are presented, for various bins in $E_{\tau}$,
in Table 5 and compared with the expected
rates for the Auger detector.  Here the 
largest Regge--inspired $\nu N$ cross section
(cf.\ Fig.\ 7) implies in most cases 
\underline{smaller} rates than our nominal
ones (based on GRV98 or CTEQ3--DIS) since 
the attenuation (absorption) exponential in
(3.3) becomes dominant at neutrino energies
above $10^9$ GeV despite the rather small
column depth involved.  In order to learn to
what extent the measured shower rates in
certain bins of $E_{\tau}$ will be able to
delineate specific features of the energy
profile of the various theoretical cosmic 
neutrino fluxes $\Phi(E_{\nu})$ in Fig.\ 1,
we presented in Table 6 the individual
$E_{\nu}$--binned contributions to the 
nominal TA--rates for given $E_{\tau}$--bins.
In any case it will be important to measure
the detailed energy ($E_{\rm sh}$ and 
$E_{\tau}$) dependencies for exploring 
extragalactic neutrino sources at highest
energies of $10^{21}$ eV.

Finally, as an illustration of possible
implications of  `new' physics, we calculated
highly speculative expectations of string
scenarios with large  `extra dimensions' at
low TeV--scales.  The additional contributions,
due to the exchange of massive `graviton' 
\mbox{spin--2}
Kaluza--Klein excitations in neutral current
neutrino--quark (gluon) scattering to the
relevant quasi--horizontal hadronic air
showers may be enormous as compared with the
moderate SM event rates for the Auger
detector and could be detected in the not too
distant future. The viability of such  `new'
physics scenarios may be tested even by the
smaller ground level AGASA and Fly's Eye
(HiRes) detectors. 
\vspace{1.5cm}

\noindent{\large\bf{Acknowledgements}}

\noindent We thank M.\ Gl\"uck for many helpful 
discussions as well as for his collaboration
at the initial stage of this work,
and P.\ Billoir for a clarifying correspondence.
We are also indebted to S.I.\ Yanush for
providing us with the computer code of the
Regge--inspired DL+CTEQ5 structure functions.
This work has been supported in part by the
`Bundesministerium f\"ur Bildung und Forschung',
Berlin/Bonn.

\newpage
%
%
\pagestyle{empty}
\setlength{\oddsidemargin}{-0.5cm}
\begin{table}[th]
\footnotesize
\parbox{17cm}
{\footnotesize{Table 1: Total upward $\mu^+$+$\mu^-$ event rates per year from $\nu_{\mu}$N and $\bar{\nu}_{\mu}$N interactions in rock calculated according to (2.1) for various muon energy thresholds E$_{\mu}^{min}$ and the appropriate cosmic neutrino fluxes in Fig.1. The numbers in parentheses refer to the event rates with no regeneration, i.e. $\Psi\equiv$1 in (2.10). The background ATM rates are shown for comparison.\newline}}
\centering 
\begin{tabular}{|c|c||c|c|c|c|c|}
\hline
\raisebox{-1.5ex}[-1.5ex]{flux} &\raisebox{-1.5ex}[-1.5ex]{detector}& \multicolumn{5}{c|}{muon-energy threshold E$_{\mu}^{min}$/GeV} \\
      &        & $10^3$& $10^4$& $10^5$ & $10^6$& $10^7$\\ \hline\hline
      & ANTARES& 498  (490) &16.92 (16.62) & 0.20 (0.20) &0.001 (0.001) & -\\ 
ATM   & AMANDA-II& 1362 (1352) &37.8 (36.8) & 0.34 (0.32) &0.0016 (0.0014)&- \\      & IceCube& 8860 (8800) & 163.4 (159)& 1.3 (1.2) &0.006 (0.006) & -\\
\hdashline\hline
      & ANTARES&   411 (359)&248 (215)& 89.3 (76.1)&13.0 (10.8)&0.53 (0.45)\\ 
AGN-SS& AMANDA-II& 699 (612)&408 (355)& 137 (116)  &19.3 (16.1)&0.79 (0.67)\\       & IceCube&  2687 (2356)&1547 (1346)&514 (439) &72.6 (60.7)&3.00 (2.53)\\ 
\hline
      & ANTARES&  13.7 (12.6)&5.00 (4.30)&1.98 (1.61)&0.90 (0.72)&0.32 (0.26)\\AGN-M95&AMANDA-II&29.1 (27.2)&8.62 (7.55)&2.98 (2.42)&1.34 (1.07)&0.46 (0.38)\\      & IceCube& 143  (135)  &33.7 (29.5)&11.2 (9.11)&5.04 (4.05)&1.74 (1.43)\\
\hline
      & ANTARES&  0.60 (0.54)&0.32 (0.28)&0.08 (0.07)&0.010 (0.008)&0.0003 (0.0003)\\ 
GRB-WB&AMANDA-II& 1.10 (1.00)&0.56 (0.50)&0.13 (0.11)&0.015 (0.012)&0.0005 (0.0004)\\  
      & IceCube&  4.35 (3.94)&2.13 (1.91)&0.49 (0.42)&0.055  (0.046)&0.0018 (0.0015)\\
\hline
      & ANTARES&  0.62 (0.51)&0.45 (0.36)&0.26 (0.21)&0.12 (0.098)& 0.045 (0.037)\\ 
TD-SLBY&AMANDA-II&0.97 (0.80)&0.68 (0.56)&0.39 (0.32)&0.18 (0.145)& 0.067 (0.054)\\  
      & IceCube&  3.70 (3.06)&2.57 (2.11)&1.47 (1.19)&0.68 (0.550)& 0.250 (0.210)\\
\hline
      & ANTARES&  0.006 (0.005)&0.005 (0.004)&0.003 (0.003)&0.002 (0.002)&0.001 (0.001)\\
TD-SLSC&AMANDA-II&0.010 (0.008)&0.007 (0.006)&0.005 (0.004)&0.003 (0.003)&0.002 (0.001)\\  
      & IceCube&  0.036 (0.030)&0.028 (0.022)&0.020 (0.015)& 0.012 (0.001)&0.006 (0.005)\\
\hline
      & ANTARES&  0.008 (0.007)& 0.007 (0.006)&0.006  (0.005)&0.005 (0.004)&0.003 (0.003)\\ 
Z-burst&AMANDA-II& 0.013 (0.010)&0.011 (0.009)& 0.009 (0.007)&0.007 (0.005)&0.005 (0.004)\\  
      & IceCube&   0.047 (0.038)&0.041 (0.032)& 0.033 (0.026)&0.026 (0.020)&0.019 (0.015)\\ 
\hline
\end{tabular}
\end{table}
\newpage
%
%
\setlength{\oddsidemargin}{0.0cm}
\begin{table}[ht]
\centering
\normalsize
\parbox{12.5cm}{\footnotesize{Table 2: Total downward $\mu^++\mu^-$ event rates per year arising from $\nu_\mu N$ and $\bar{\nu}_\mu N$ interactions in a detector with an effective volume $V_{eff}=A_{eff}\times 1$km $=1$km$^3$ of water calculated according to (2.3) with $S\equiv 1$ for various muon energy thresholds $E_\mu^{min}$. The background ATM rates are shown for comparison.\newline}}
\begin{tabular}{|c||c|c|c|c|c|c|}
\hline
     & \multicolumn{6}{c|}{$E_\mu^{min}$ [GeV]}               \\
\raisebox{1.5ex}{flux} & $10^5$ & $10^6$ & $10^7$ & $10^8$ & $10^9$ & $10^{10}$ \\
\hline\hline
ATM     & 5.08 & 0.041  & 3$\times 10^{-4}$ &  -  & -  &-  \\
\hdashline
AGN-SS  & 510  & 207    & 30.9   & 0.34   &$1.0\times 10^{-4}$&  -       \\
AGN-M95 & 11.8 & 8.95   & 7.09   & 3.74   & 0.88   & 0.054    \\
GRB-WB  & 0.61 & 0.16   & 0.02   &$5\times 10^{-4}$ &-&- \\
TD-SLBY & 1.47 & 1.30   & 1.00   & 0.63   & 0.30   & 0.10  \\
TD-SLSC &0.028 & 0.027  & 0.025  & 0.022  & 0.017  & 0.010    \\
Z-burst & 0.082& 0.082  & 0.082  &0.081   & 0.077  & 0.063    \\
\hline
\end{tabular}
\end{table}
%
%
\begin{table}[ht]
\centering
\normalsize
\bigskip
\parbox{12.5cm}{\footnotesize{Table 3: Total downward $\mu^++\mu^-$ event rates per year for a detector with $V_{eff}=1$km$^3$ as in Table 2 but using the DL+CTEQ5 structure functions [62] with their Regge model inspired small-x extrapolation [61]. The background ATM rates are shown for comparison.\newline}}
\begin{tabular}{|c||c|c|c|c|c|c|}
\hline
& \multicolumn{6}{c|}{$E_\mu^{min}$ [GeV]}               \\
\raisebox{1.5ex}{flux} & $10^5$ & $10^6$ & $10^7$ & $10^8$ & $10^9$ & $10^{10}$ \\
\hline\hline
ATM     & 5.69 & 0.047  & 3$\times 10^{-4}$ &  -  & - &-   \\
\hdashline
AGN-SS  & 586  & 241    & 35.9   & 0.40   &$1.0\times 10^{-4}$& -        \\
AGN-M95 & 13.8 & 10.5   & 8.35   & 4.45   & 1.09   & 0.072    \\
GRB-WB  &0.70  & 0.19   & 0.02   &$6\times 10^{-4}$&-&-   \\
TD-SLBY &1.75  & 1.55   & 1.21   & 0.77   & 0.38   & 0.13     \\
TD-SLSC &0.036 & 0.035  & 0.033  & 0.029  & 0.023  & 0.015    \\
Z-burst &0.116 & 0.115  & 0.115  & 0.114  & 0.109  & 0.091    \\
\hline
\end{tabular}
\end{table}
\newpage
%
%
\setlength{\oddsidemargin}{-1.0cm}
\begin{table}[th]
\footnotesize
\centering
\parbox{18.2cm}
{\footnotesize{Table 4: Total annual event rates for the Pierre Auger Observatory for horizontal air showers induced by ($\nu_e+\bar{\nu}_e$)N and ($\nu_{\mu}+\bar{\nu}_{\mu}$)N CC and NC interactions calculated according to (3.1) using the geometrical acceptances of Fig.8 and the appropriate cosmic neutrino fluxes of Fig.1. The numbers in parentheses refer to the rates as obtained by using the (optimal) DL+CTEQ5 structure functions [62] with their Regge model inspired small-x extrapolation [61]. For comparison the expected rates for the Telescope Array are also displayed, where available, which have been estimated [26] using CTEQ4-DIS parton distributions and assuming a duty cycle of 0.1 for observing the atmospheric fluorescence light.\newline}}
\begin{tabular}{|c|c|c c:c||c c:c|}
\cline{3-8}
 \multicolumn{2}{c}{}&  \multicolumn{6}{|c|}{detectors} \\
 \multicolumn{2}{c}{}& \multicolumn{3}{|c||}{Auger} & \multicolumn{3}{c|}{TA} \\ \hline
flux & E$_{sh}$(GeV)& CC($\nu_e,\nu_{\mu}$) & NC & total &CC($\nu_e,\nu_{\mu}$)& NC & total \\ \hline
      &$10^8-10^9$      & 0.032 (0.039)  & 0.007 (0.008)&0.039 (0.047) & & &\\ 
      &$10^{9}-10^{10}$ & 0.0001 (0.0001)& 0.000008 (0.000009) &0.0001 (0.0001) & & & \\    
AGN-SS&$10^{10}-10^{11}$& $-$ ($-$)  &$-$ ($-$) &$-$ ($-$) &\multicolumn{2}{c}{$-$} &\\ 
      &$10^{11}-10^{12}$& $-$ ($-$)  &$-$ ($-$)  & $-$ ($-$) & & &\\\cdashline{2-8}
      & total           & 0.032 (0.039)  &0.007 (0.008) &0.039 (0.047) & $-$ &$-$ &$-$\\
\hline
       &$10^8-10^9$      & 1.42 (1.78)& 0.57 (0.64)&1.99 (2.42)  & & & \\ 
       &$10^{9}-10^{10}$ & 1.52 (1.97)& 0.27 (0.33)&1.79 (2.30)& & &\\    
AGN-M95&$10^{10}-10^{11}$& 0.19 (0.27)& 0.015 (0.02)&0.21 (0.29) &\multicolumn{2}{c}{$-$} &\\ 
       &$10^{11}-10^{12}$&$-$ ($-$)&$-$ ($-$) &$-$& & &\\ \cdashline{2-8}
       & total           &3.13 (4.02)  &0.86 (0.99) &3.99 (5.01)&9.2 &3.1 &12.3\\
\hline
       &$10^8-10^9$      &0.00011 (0.00014)&0.000031 (0.000034)&0.0001 (0.00014)& & &\\ 
       &$10^{9}-10^{10}$&$-$ ($-$)& $-$ ($-$)& $-$ ($-$) & & &\\    
GRB-WB&$10^{10}-10^{11}$&$-$ ($-$)& $-$ ($-$)& $-$ ($-$) &\multicolumn{2}{c}{$-$} &\\ 
       &$10^{11}-10^{12}$&$-$ ($-$)&$-$ ($-$)& $-$ ($-$) & & &\\ \cdashline{2-8}
       & total           &0.00013 (0.0002)&0.000034 (0.000038) &0.00016 (0.00024) &$-$ &$-$ & $-$\\\cdashline{2-8}
\hline
       &$10^8-10^9$      &0.23 (0.29)&0.11 (0.13) &0.34 (0.42)& & &\\ 
       &$10^{9}-10^{10}$ &0.49 (0.64)&0.14 (0.17) &0.63 (0.81)& & &\\    
TD-SLBY&$10^{10}-10^{11}$&0.38 (0.54)&0.07 (0.10) &0.45 (0.64)&\multicolumn{2}{c}{$-$} &\\ 
       &$10^{11}-10^{12}$&0.13 (0.22)&0.02 (0.02) &0.15 (0.24)& & &\\ \cdashline{2-8}
       & total           &1.23 (1.69)&0.34 (0.42)&1.57 (2.11)&$-$ & $-$&$-$\\
\hline
       &$10^8-10^9$      &0.006 (0.007)&0.004 (0.004)&0.010 (0.011) & & &\\ 
       &$10^{9}-10^{10}$ &0.021 (0.028)&0.009 (0.011)&0.030 (0.039)& & &\\    
TD-SLSC&$10^{10}-10^{11}$&0.035 (0.052)&0.010 (0.014)&0.045 (0.066)&\multicolumn{2}{c}{$-$} &\\ 
       &$10^{11}-10^{12}$&0.037 (0.060)&0.005 (0.008)&0.042 (0.068) & & &\\ \cdashline{2-8}
       & total           &0.099 (0.147)& 0.028 (0.037)&0.127 (0.184) &0.15 &0.07 &0.22\\
\hline
       &$10^8-10^9$      &0.014 (0.017)&0.011 (0.012)&0.025 (0.029) & & &\\ 
       &$10^{9}-10^{10}$ &0.074 (0.099)&0.040 (0.049)&0.114 (0.148)& & &\\    
Z-burst&$10^{10}-10^{11}$&0.188 (0.277)&0.067 (0.093)&0.255 (0.370)&\multicolumn{2}{c}{$-$} &\\ 
       &$10^{11}-10^{12}$&0.315 (0.519)&0.049 (0.078)&0.364 (0.597)& & &\\ \cdashline{2-8}
       & total           &0.59 (0.91) &0.17 (0.23) &0.76 (1.14) &0.79 &0.40 &1.19\\
\hline
\end{tabular}
\end{table}
\newpage
%
%
\setlength{\oddsidemargin}{0.0cm}
\begin{table}[th]
\centering
\normalsize
\parbox{12.3cm}
{\footnotesize{Table 5: Annual nominal event rates produced by Earth-skimming $\nu_{\tau}$'s for the fluorescence Telescope Array detector (10 stations) calculated according to (3.2) with D=0.1 and the aperture for the detections of $\tau$ leptons through their dominant decay to electromagnetic showers taken from [25]. The numbers in parentheses are the rates resulting from the largest Regge-inspired $\nu$N cross section in Fig.7. The estimated total TA rates of [26] are also displayed, where available. The total Auger rates are obtained from folding the effective Auger aperture [43]  ('BS+PP+DIS low' curve in Fig.9 of [43]) with the appropriate cosmic $\nu_{\tau}$-fluxes in Fig.1. These latter event rates refer to the Auger ground array (not the fluorescence detector) with a duty cycle of 1.\newline}}
\begin{tabular}{|c|c|c|c|c|}
\cline{3-5}
 \multicolumn{2}{c}{}&  \multicolumn{3}{|c|}{detectors} \\
\cline{1-2}
flux & E$_{\tau}$(GeV)& TA & TA([26]) &Auger([43]) \\ \hline
      &$10^8-10^9$      &1.07 (1.12) & & \\ 
      &$10^{9}-10^{10}$ &0.0006 (0.0006) & & \\    
AGN-SS&$10^{10}-10^{11}$&$-$ ($-$)  &$-$ & $-$    \\ 
      &$10^{11}-10^{12}$&$-$  ($-$) & & \\\cdashline{2-5}
      & total           &1.07 (1.12)&$-$ & 1.15 \\
\hline
       &$10^8-10^9$      &47.4 (46.1) & & \\ 
       &$10^{9}-10^{10}$ &14.8 (12.8) & &\\    
AGN-M95&$10^{10}-10^{11}$&0.25 (0.19) &$-$ & $-$\\ 
       &$10^{11}-10^{12}$&$-$  ($-$)& &\\ \cdashline{2-5}
       & total           &62.5 (59.1)&42.8 & 72.8\\
\hline
       &$10^8-10^9$      &0.0035 (0.0036)& &\\ 
       &$10^{9}-10^{10}$ &0.0002 (0.0002)& &\\    
GRB-WB&$10^{10}-10^{11}$ &$-$ ($-$)&$-$ & $-$\\ 
       &$10^{11}-10^{12}$&$-$ ($-$)& &\\ \cdashline{2-5}
       & total           &0.0037 (0.0038)&$-$ & 0.0039\\\cdashline{2-5}
\hline
       &$10^8-10^9$      &7.62 (7.22)& &\\ 
       &$10^{9}-10^{10}$ &5.21 (4.35)& &\\    
TD-SLBY&$10^{10}-10^{11}$&0.54 (0.40)&$-$ & $-$\\ 
       &$10^{11}-10^{12}$&0.008 (0.005) & &\\ \cdashline{2-5}
       & total           &13.4 (11.98)&$-$ & 17.0\\
\hline
       &$10^8-10^9$      &0.19 (0.17)& &\\ 
       &$10^{9}-10^{10}$ &0.23 (0.18)& &\\    
TD-SLSC&$10^{10}-10^{11}$&0.05 (0.04)&$-$ &$-$\\ 
       &$10^{11}-10^{12}$&0.002 (0.001) & &\\ \cdashline{2-5}
       & total           &0.47 (0.39)&0.34 & 0.66\\
\hline
       &$10^8-10^9$      &0.34 (0.29)& &\\ 
       &$10^{9}-10^{10}$ &0.80 (0.63)& &\\    
Z-burst&$10^{10}-10^{11}$&0.27 (0.19)&$-$ & $-$\\ 
       &$10^{11}-10^{12}$&0.017 (0.011)& &\\ \cdashline{2-5}
       & total           &1.43 (1.12)&1.19 & 2.18\\
\hline
\end{tabular}
\end{table}
\newpage
%
%
\begin{table}[ht]
\centering
\normalsize
\parbox{11cm}{\footnotesize{Table 6: Contributions to the nominal TA event rates for given $E_\tau$-bins in Table 5 from various increasing $E_\nu$-bins of three representative cosmic neutrino fluxes.\newline}}
\begin{tabular}{|c:c|c|c|c|c|}
\hline
\multicolumn{2}{|c|}{flux} & AGN-M95 & TD-SLBY & Z-burst  \\
\hline
$E_\tau$ [GeV] & $E_\nu$ [GeV] & TA & TA & TA \\
\hline
$10^8-10^9$ & $10^8-10^9$    & 32.11 & 4.05 & 0.055 \\
            & $10^9-10^{10}$ & 15.04 & 3.27 & 0.19 \\
            & $10^{10}-10^{11}$ & 0.26 & 0.28 & 0.082 \\
            & $10^{11}-10^{12}$ & $-$ & 0.012 & 0.017 \\
\hdashline
\multicolumn{2}{|c|}{total} & 47.4 & 7.6 & 0.34 \\
\hline
$10^9-10^{10}$ & $10^9-10^{10}$ & 13.59 & 3.87 & 0.35 \\
               & $10^{10}-10^{11}$ & 1.17 & 1.28 & 0.37 \\
               & $10^{11}-10^{12}$ & 0.0001 & 0.054 & 0.08 \\
\hdashline
\multicolumn{2}{|c|}{total} & 14.8 & 5.2 & 0.8 \\
\hline
$10^{10}-10^{11}$ & $10^{10}-10^{11}$ & 0.25 & 0.48 & 0.18 \\
               & $10^{11}-10^{12}$ & 0.0002 & 0.062 & 0.09 \\
\hdashline
\multicolumn{2}{|c|}{total} & 0.25 & 0.54 & 0.27 \\
\hline
$10^{11}-10^{12}$ & $10^{11}-10^{12}$ & $-$ & 0.008 & 0.017 \\
\hline
\end{tabular}
\end{table}
\bigskip
%
%
\begin{table}[th]
\normalsize
\centering
\parbox{12.0cm}{\footnotesize{Table 7: Total annual event rates for the Pierre Auger Observatory for horizontal air showers as implied by the SM and the additional contributions due to the exchange of KK gravitons in a low-scale scenario with large 'extra dimensions'. The rates are calculated according to (3.1), with the additional graviton KK exchange contribution given in (4.5), and using the geometrical acceptances of Fig.8. The first entry corresponds to using the unitarity interpolation (4.3) (dotted curve in Fig.10) and the second entry is the result of using (4.4) (dashed curve in Fig.10).\newline}}
\begin{tabular}{|c||c|c|c|c|}
\hline
     & \multicolumn{4}{c|}{$E_{sh}$ threshold $E_{th}$/GeV}      \\
\raisebox{1.5ex}{flux} & $10^8$ & $10^9$ & $10^{10}$ & $10^{11}$ \\
\hline\hline
AGN-SS  & 0.54/0.38   & 0.001/0.001  & -/-       & -/- \\
AGN-M95 & 401/880     & 151/398      & 7.2/27.9  & 0.001/0.007 \\
GRB-WB  & 0.007/0.011 & 0.001/0.004  & -/0.001   & -/- \\
TD-SLBY & 352/2996    & 267/2487     & 118/1350  & 19.8/289 \\
TD-SLSC & 47.0/703    & 42.4/624     & 27.5/407  & 7.3/114 \\
Z-burst & 342/6305    & 321/5696     & 233/3919  &  76.2/1250 \\
\hline
\end{tabular}
\end{table}
\newpage
%
%
%
\setlength{\topmargin}{-1.5cm}
\begin{figure}[th]
\centering
\includegraphics[width=9cm,angle=270]{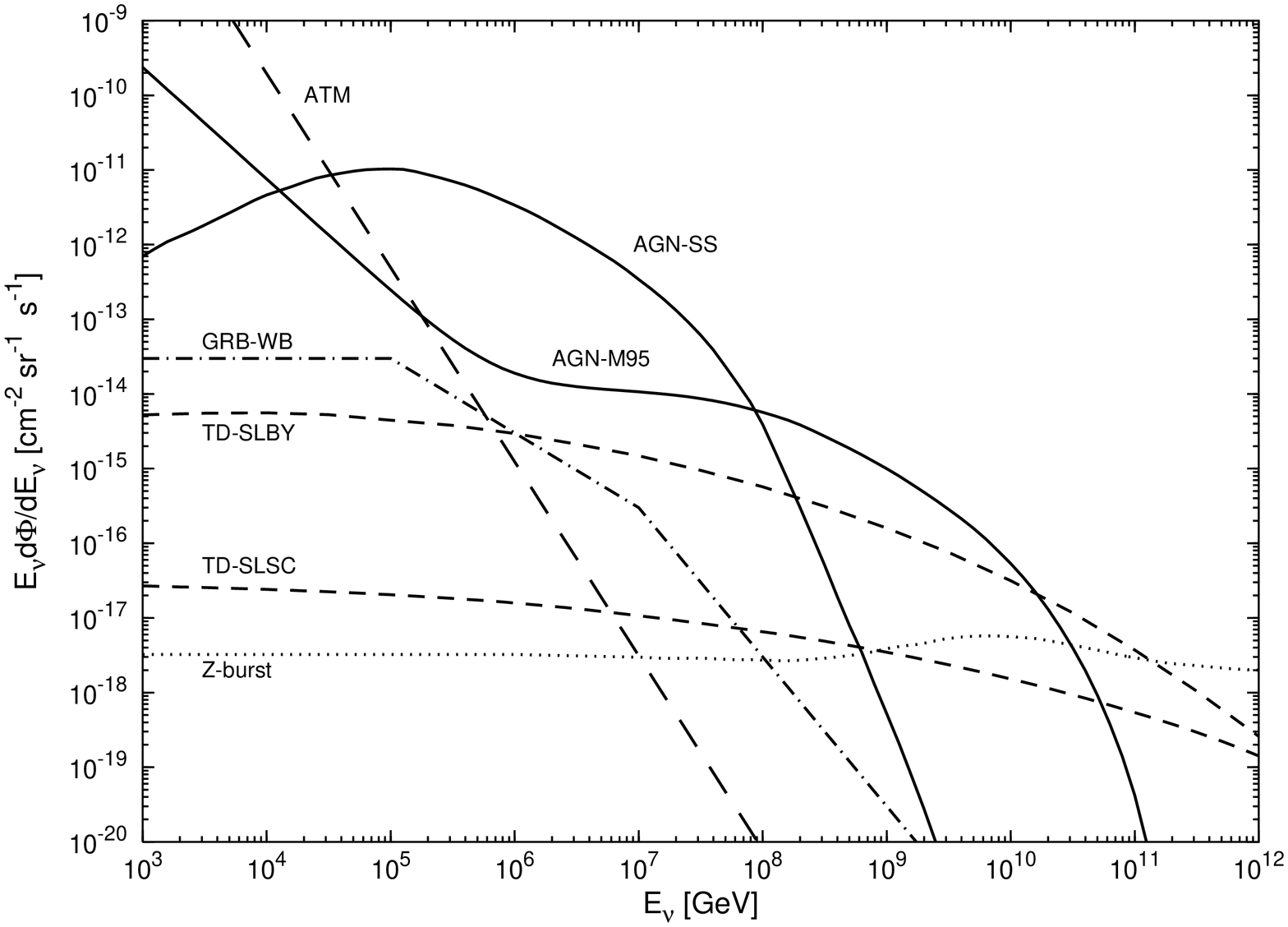}
\parbox{12cm}{
\medskip\footnotesize{Figure 1: Representive differential fluxes of muon neutrinos ($\nu_{\mu}$+$\bar{\nu}_{\mu}$) from active galactic nuclei (AGN-SS [9] and AGN-M95 [8]), gamma ray bursts (GRB-WB [11]), topological defects (TD-SLSC [14] and TD-SLBY [15]) and Z-bursts [18]. Due to naive channel counting in pion production and decay at the production site ($\nu_e:\nu_{\mu}=1:2$) and  maximal mixing, $\nu_e:\nu_{\mu}:\nu_{\tau}=1:1:1$, these fluxes are divided equally between $\mu$, $e$ and $\tau$ neutrinos when they reach Earth (i.e. will be devided by a factor of 2). The background angle-averaged atmospheric (ATM) neutrino $\nu_\mu+\bar{\nu}_\mu$ flux [19,20,21] is shown for illustration by the long-dashed curve.}\newline}
%
%
\centering
\includegraphics[width=9cm,angle=270]{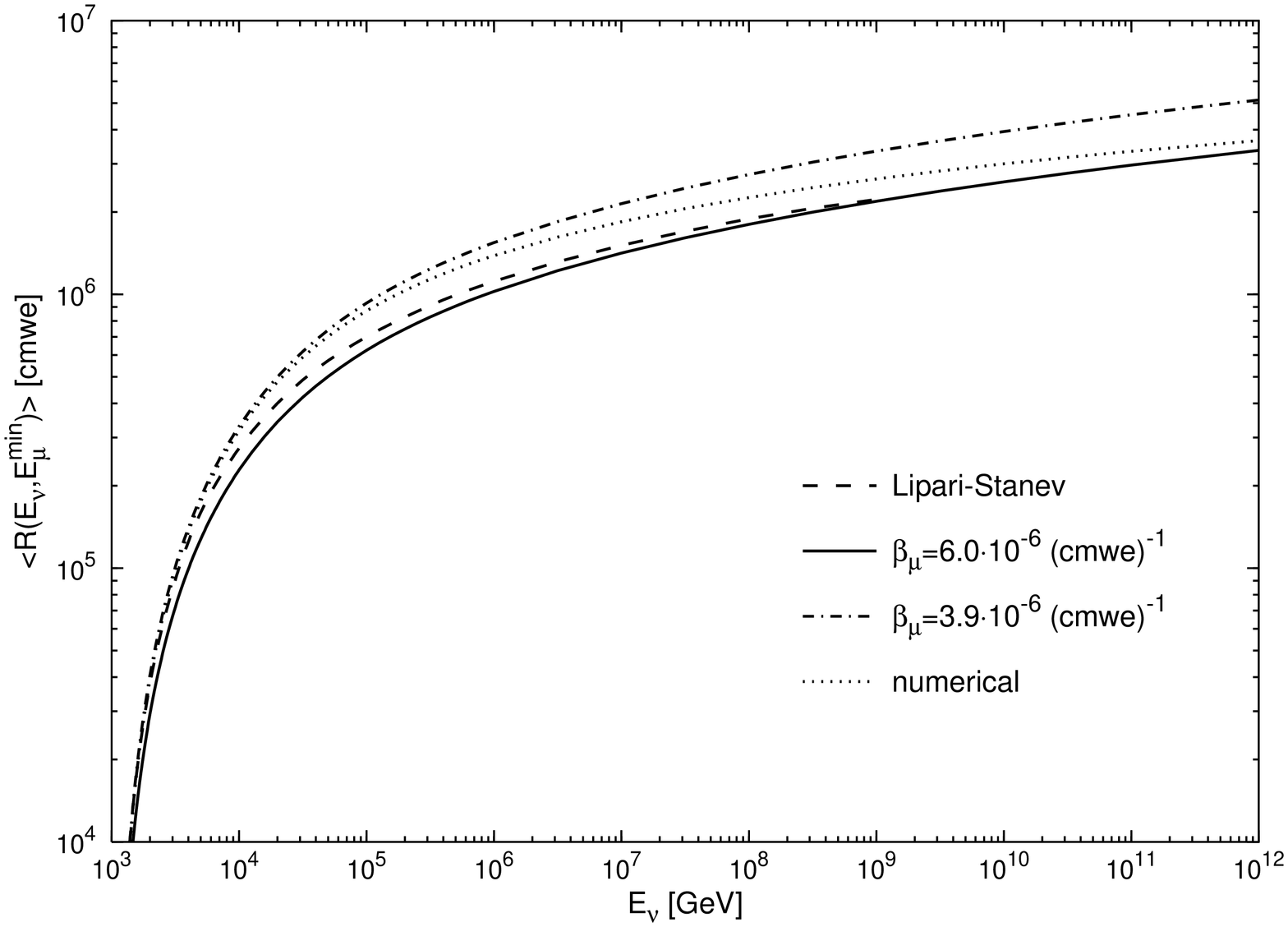}
\parbox{12cm}{
\medskip\footnotesize{Figure 2: Mean ranges in rock of muons produced in CC interactions of neutrinos with energies E$_{\nu}$ according to (2.5) with E$_{\mu}^{min}=$1TeV. The analytic ranges follow from (2.7) for constant values of $\alpha_{\mu}$ and $\beta_{\mu}$, and the numerical range follows from (2.6) using (2.8). The Lipari-Stanev range ist taken from [48] which extends only up to 10$^9$ GeV.}}
\end{figure}
\newpage
%
%
\begin{figure}[th]
\centering
\includegraphics[width=12cm]{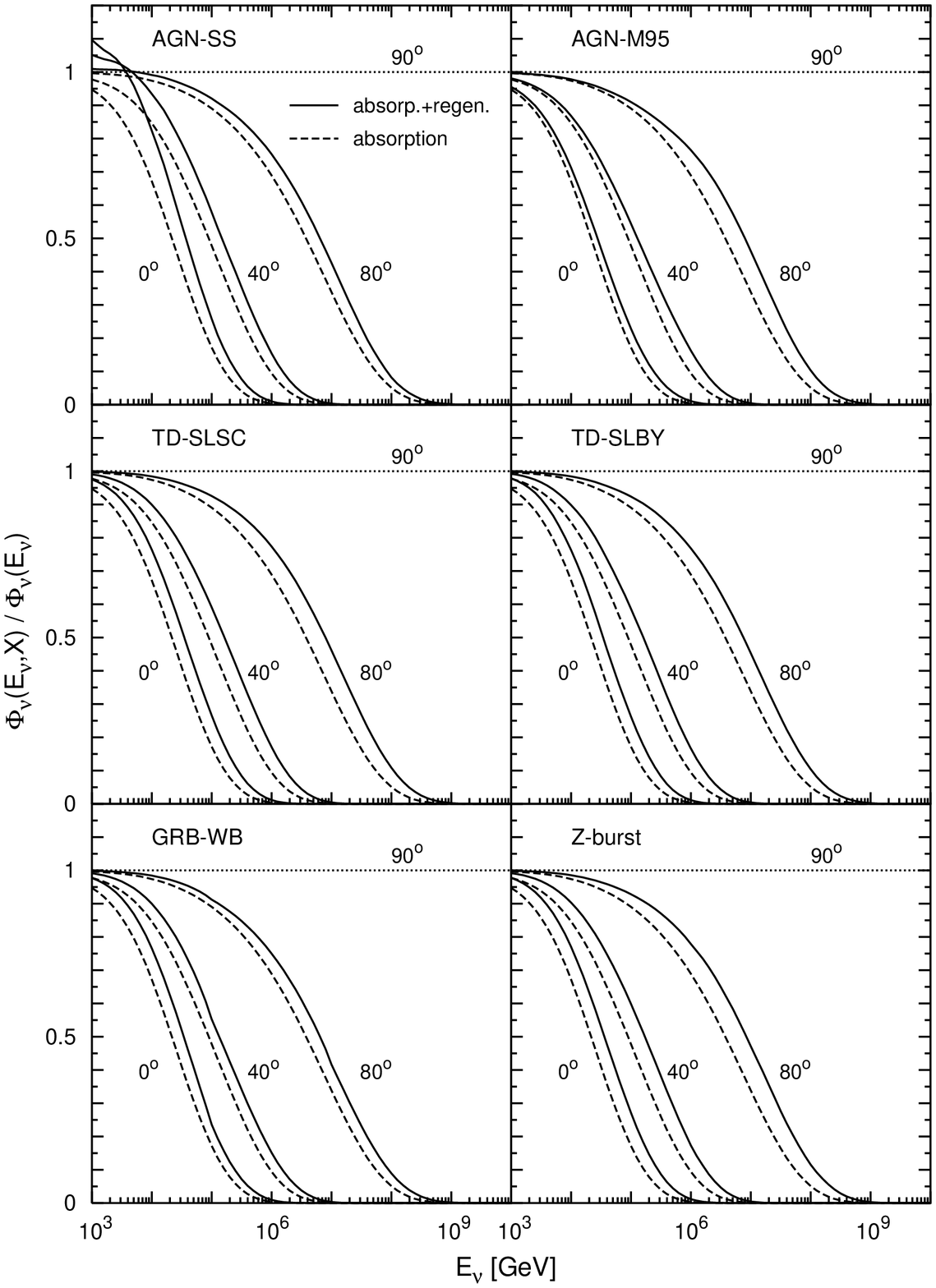}
\parbox{12cm}{\footnotesize{Figure 3(a): The differential shadow factor in (2.10) for three different nadir angles of cosmic neutrino fluxes incident at the Earth's surface with an initial flux $\Phi_\nu(E_\nu,X=0)\equiv\Phi_\nu(E_\nu)=\Phi(E_\nu)/4$ with the original total $\nu_\mu+\bar{\nu}_\mu$ flux $\Phi(E_\nu)$ being given in Fig.1. The dashed curves describe the attenuation just due to absorption where regeneration is omitted, $\Psi\equiv 1$ in (2.10).}}
\end{figure}
%
%
\begin{figure}[th]
\centering
\includegraphics[width=12cm]{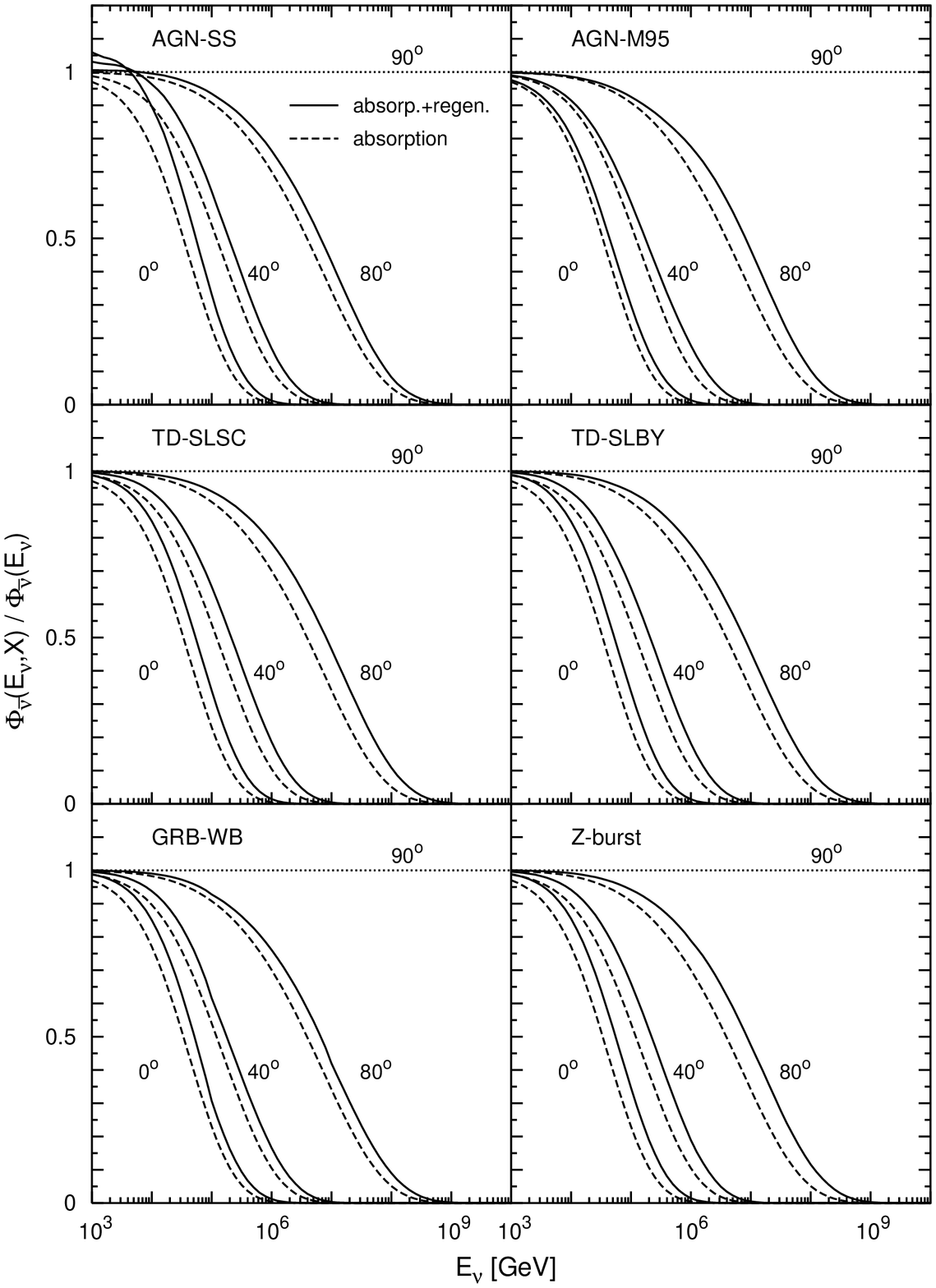}
\parbox{12cm}{\footnotesize{Figure 3(b): As in Fig. 3(a) but for antineutrinos where the initial flux at the Earth's surface is again given by $\Phi_{\bar{\nu}}(E_\nu,X=0)\equiv\Phi_{\bar{\nu}}(E_\nu)=\Phi(E_\nu)/4$ with $\Phi$ being given in Fig.1.}}
\end{figure}
%
%
\begin{figure}[th]
\centering
\includegraphics[width=12cm]{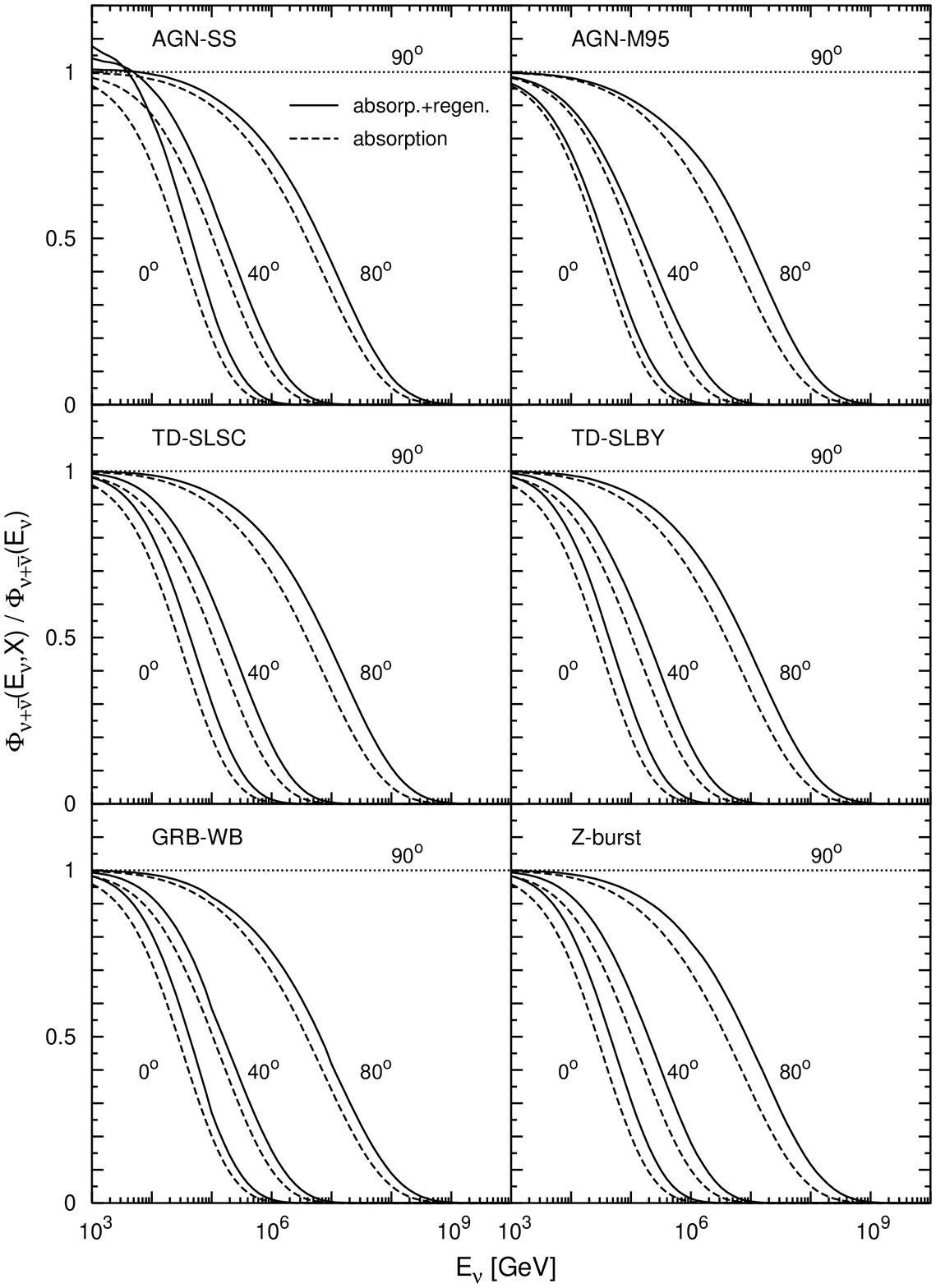}
\parbox{12cm}{\footnotesize{Figure 3(c): As in Fig. 3(a) but for the total neutrino plus antineutrino fluxes where the initial total flux at the Earth's surface is given by $\Phi_{\nu+\bar{\nu}}(E_\nu,X=0)\equiv\Phi_{\nu+\bar{\nu}}(E_\nu)=\Phi(E_\nu)/2$ with the original $\Phi$ given in Fig.1.}}
\end{figure}
\newpage
%
%
\begin{figure}[th]
\centering
\includegraphics[width=13.5cm,angle=0]{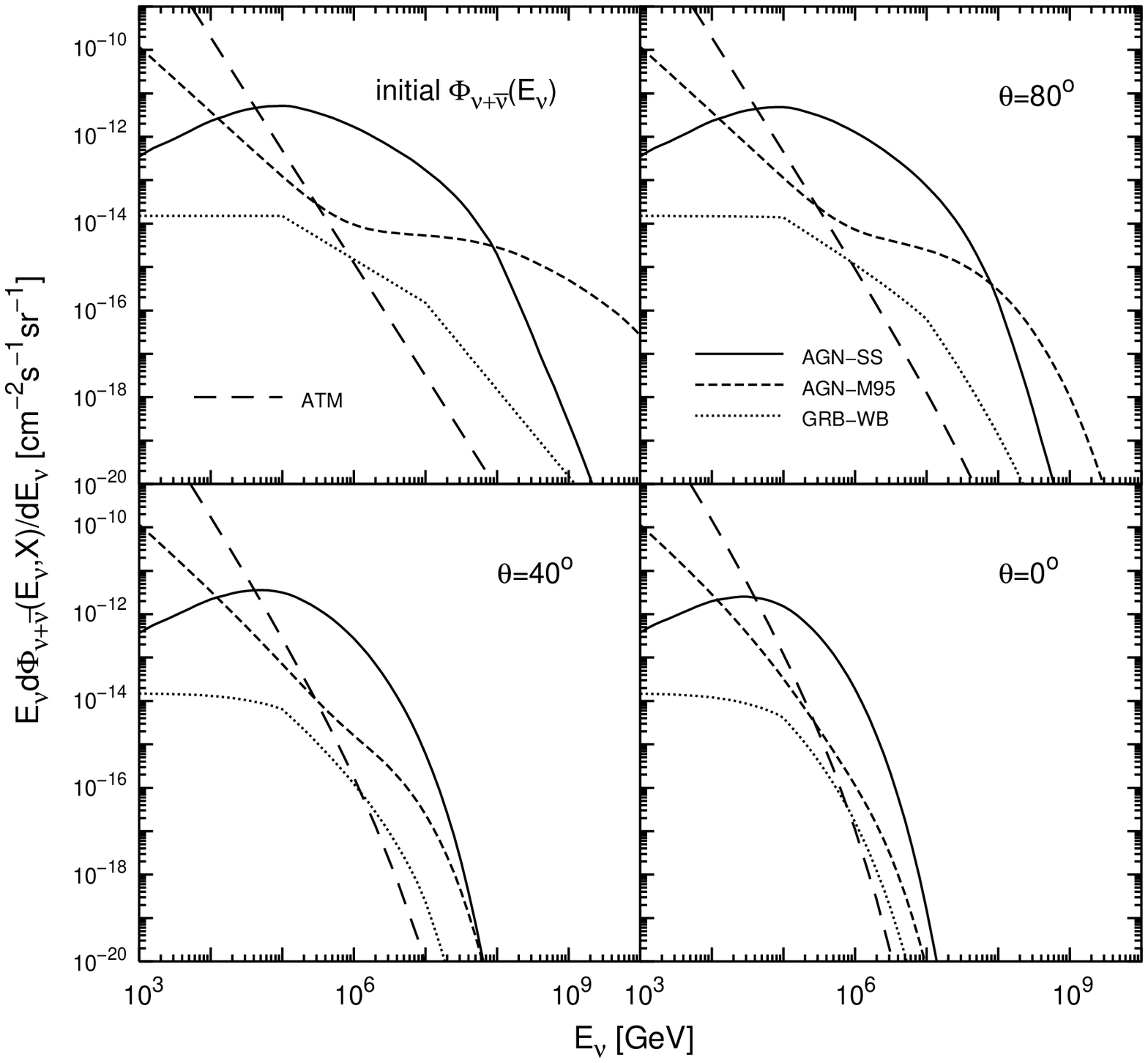}
\parbox{13.5cm}{
\medskip \footnotesize{Figure 4(a): The initial $\nu_{\mu}$+$\bar{\nu}_{\mu}$ flux at the Earth's surface $\Phi_{\nu+\bar{\nu}}$(E$_{\nu}$,X=0)$\equiv\Phi_{\nu+\bar{\nu}}$(E$_{\nu}$)=$\Phi$(E$_{\nu}$)/2, with $\Phi$(E$_{\nu}$) in Fig.1, and the flux at the detector $\Phi_{\nu+\bar{\nu}}$(E$_{\nu}$,X) for three different nadir angles corresponding to two models for AGN neutrinos, AGN-SS [9] and AGN-M95 [8], and to neutrinos from gamma ray bursts, GRB-WB [11]. The division by a factor of 2 of the original cosmic fluxes $\Phi$(E$_{\nu}$) in Fig.1 is due to maximal mixing. The background atmospheric (ATM) neutrino flux is also shown for comparison.}}
\end{figure}
%
%
\begin{figure}[th]
\centering
\includegraphics[width=13.5cm,angle=0]{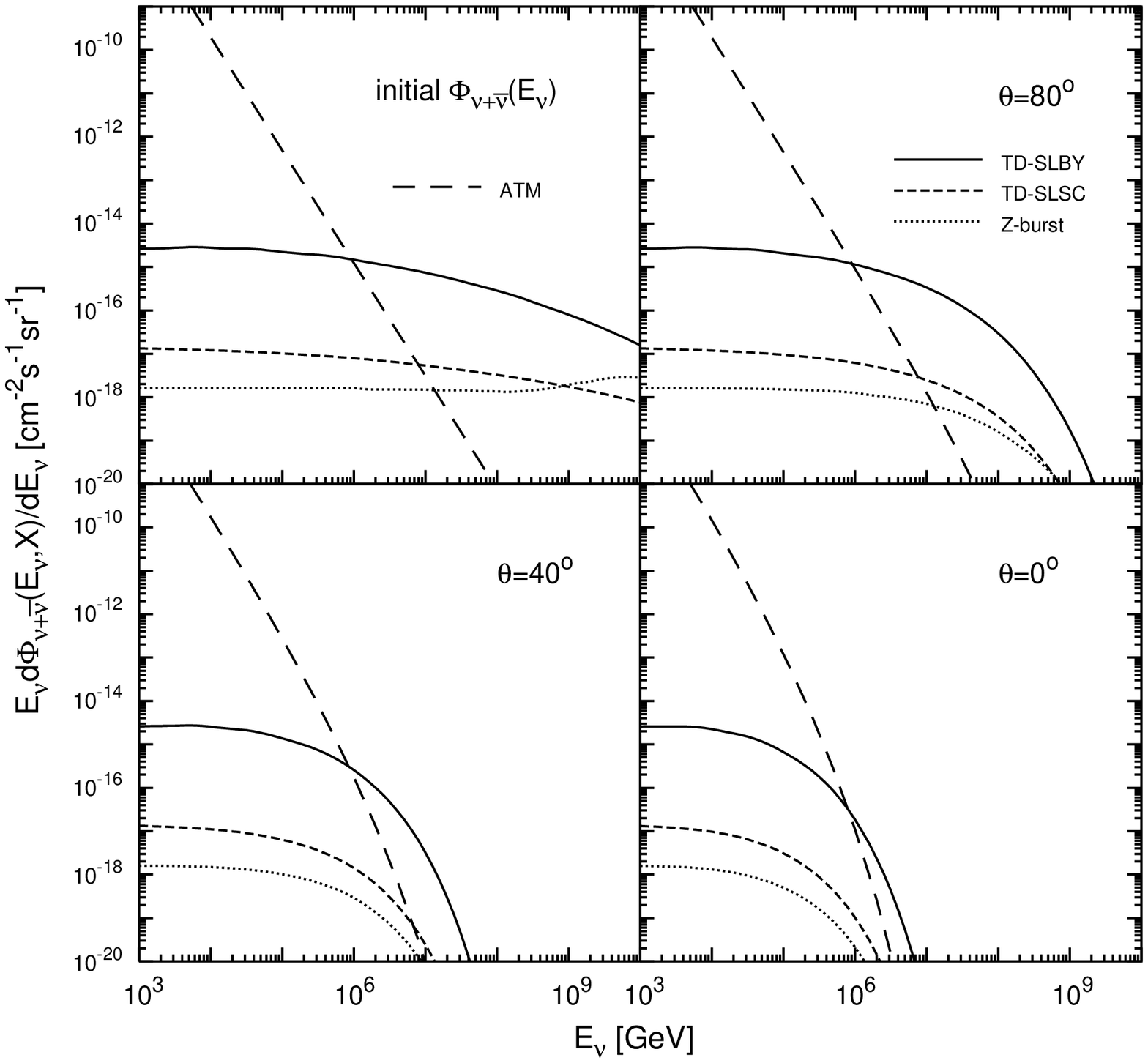}
\parbox{13.5cm}{
\medskip \footnotesize{Figure 4(b): As in Fig.4(a), but showing neutrino fluxes from two topological defects models, TD-SLBY [15] and TD-SLSC [14], and from Z-bursts [18].}}
\end{figure}
\newpage
%
%
\setcounter{figure}{4}
\begin{figure}[th]
\centering
\includegraphics[width=9cm,angle=270]{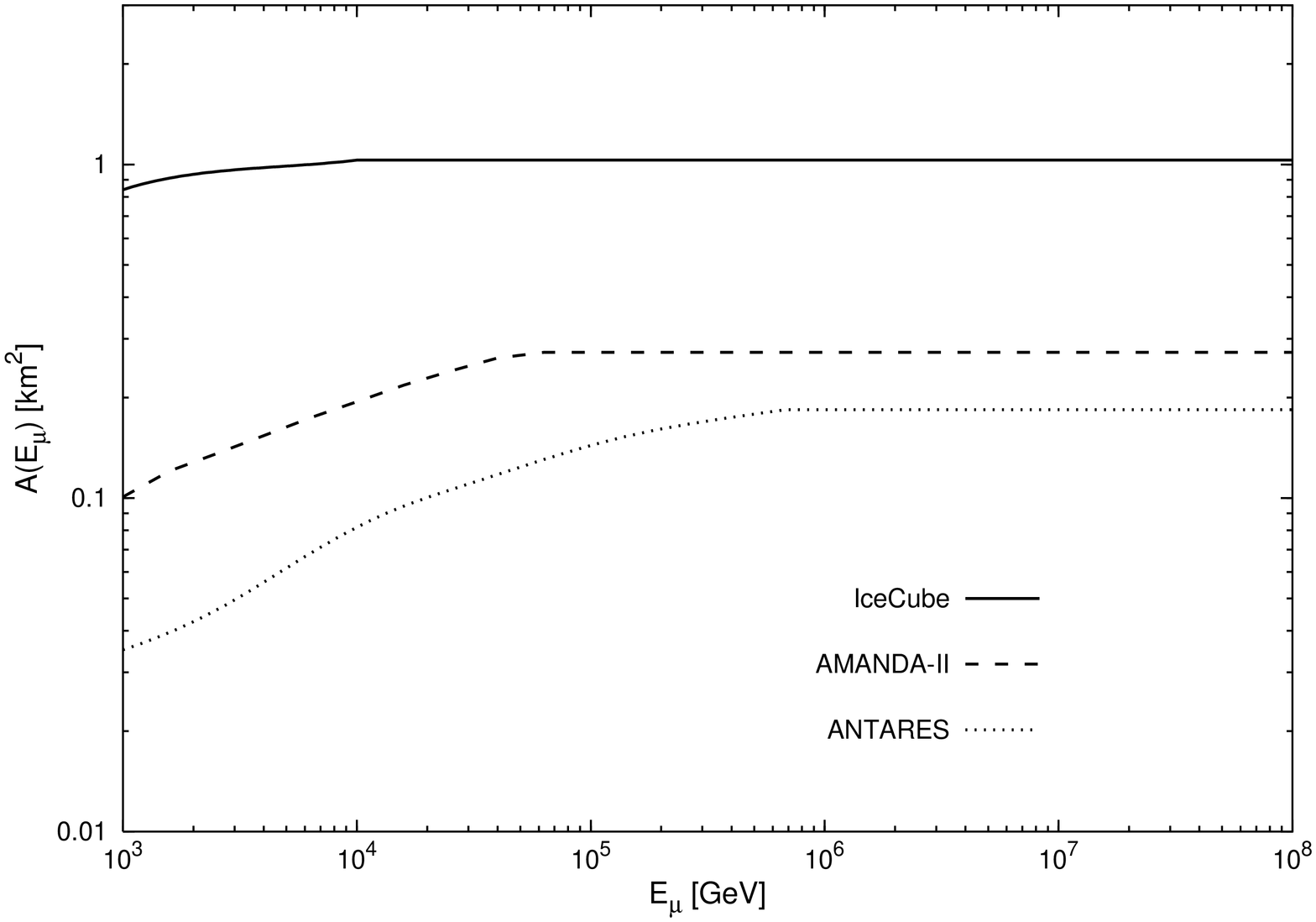}
\parbox{12cm}{
\medskip\footnotesize{Figure 5: Effective energy dependent areas for the underground detectors ANTARES [39], Amanda-II [50] and IceCube [51] used for our calculations of total event rates.\newline}}
%
%
\includegraphics[width=9cm,angle=270]{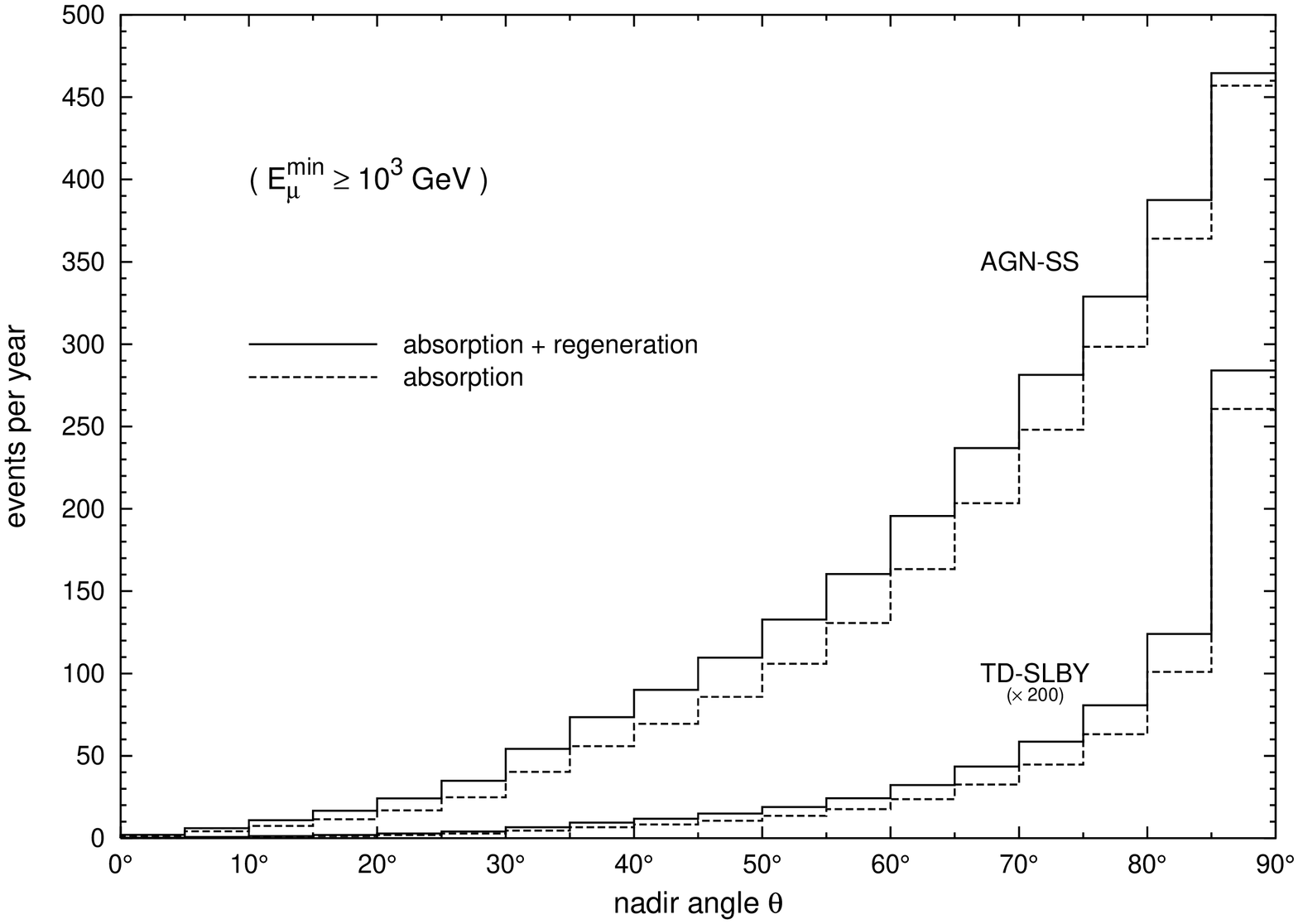}
\parbox{12cm}{
\medskip\footnotesize{Figure 6: Nadir angular dependence of upward $\mu^++\mu^-$ event rates per year for two representative neutrino flux models in Fig.1. The dashed histograms refer to events where the neutrino attenuation is caused just by absorption ($\Psi\equiv 1$ in (2.12)) with no regeneration. For definiteness an effective energy-independent detector area of $A_{eff}=1$ km$^2$ has been used in (2.1) (cf. Fig.5). The TD-SLBY rates have been multiplied by a factor of 200.}}
\end{figure}
\newpage
%
%
\begin{figure}[th]
\centering
\includegraphics[width=9cm,angle=270]{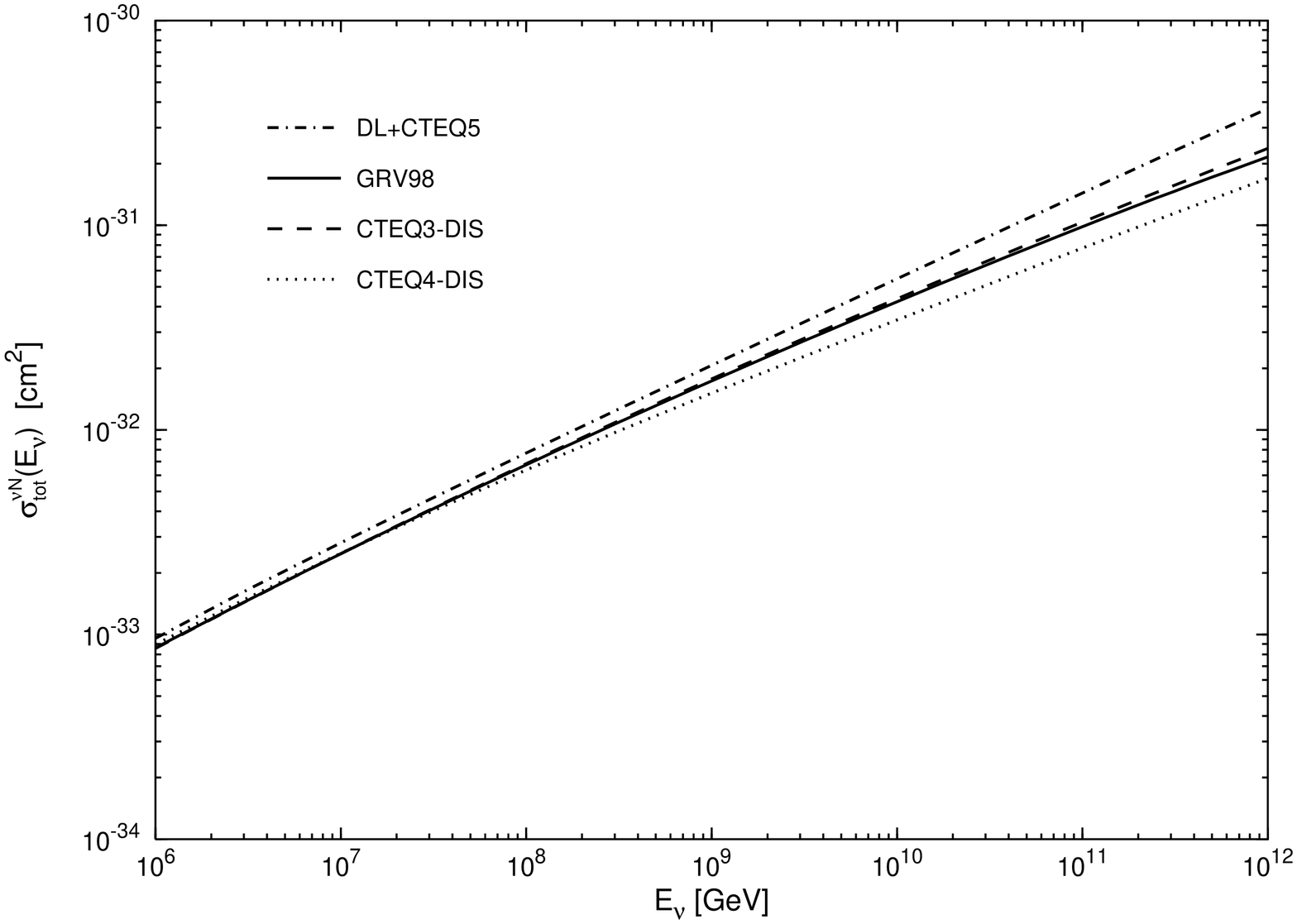}
\parbox{12cm}{
\medskip\footnotesize{Figure 7: Total (CC+NC) $\nu$N cross sections. The dashed-dotted curve refers to the cross section calculated using the DL+CTEQ5 structure functions [62] with their Regge model inspired small-x extrapolation [61]. Our nominal results are based on GRV98 [32], or equivalently on the CTEQ3-DIS [37] parton distributions. For illustration the results based on CTEQ4-DIS [38] are shown as well.\newline}}
%
%
\psfrag{ylabel}{\small $\mathcal{A}$(E$_{sh}$) [km$^3$ we sr]}
\psfrag{xlabel}{\small E$_{sh}$ [GeV]}
\includegraphics[width=9cm,angle=270]{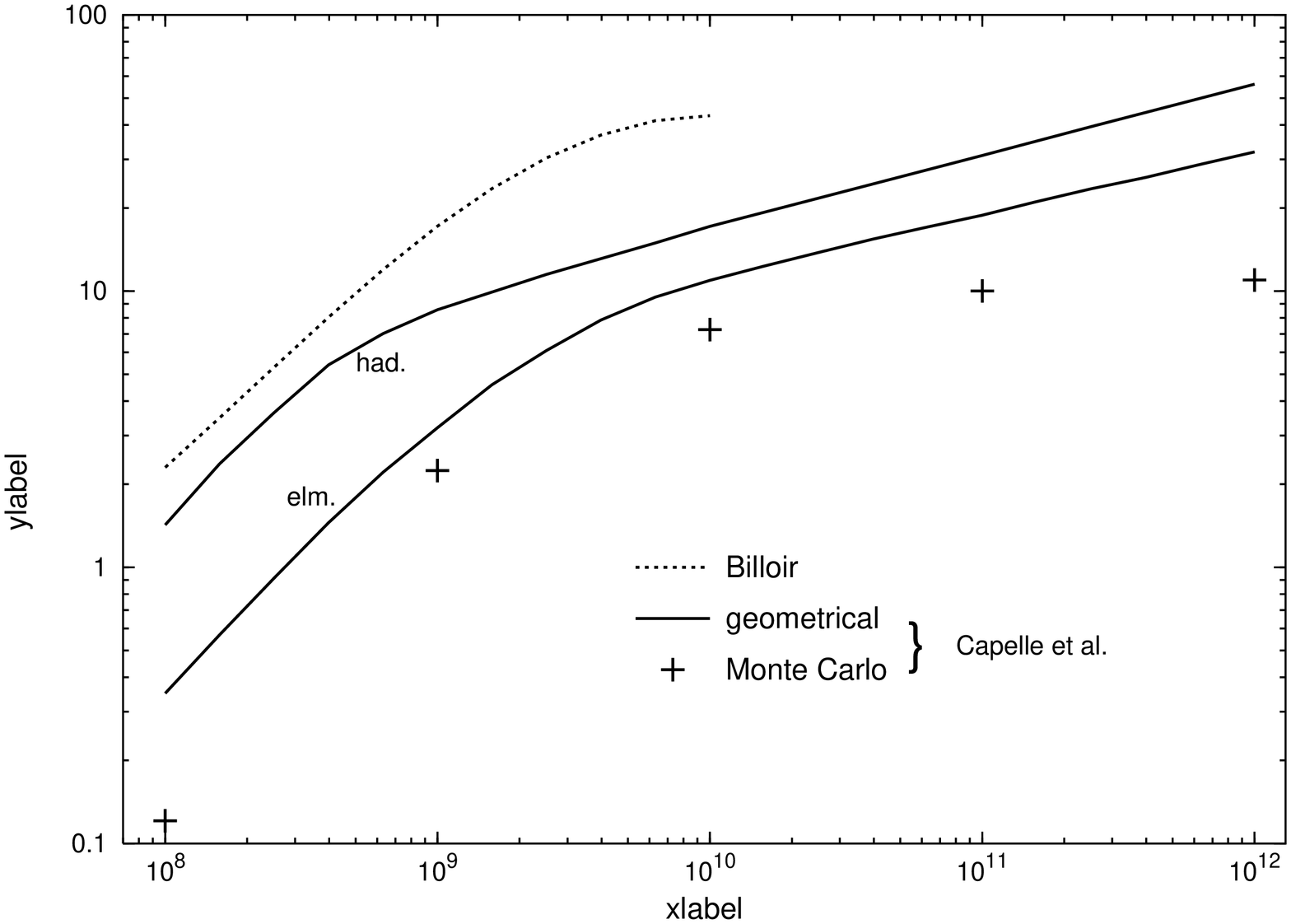}
\parbox{12cm}{
\medskip\footnotesize{Figure 8: Acceptances versus shower energy needed for calculating near-horizontal air shower event rates according to (3.1) for the Auger detector. The solid curves correspond to a geometrical integration for electromagnetic and hadronic showers of Capelle et al. [28] and the crosses are the results of a Monte Carlo simulation of all showers [28] which can be considered as a lower bound of the acceptance. For comparison the acceptance for all showers as estimated by Billoir [65] is shown by the dotted curve.}}
\end{figure}
\newpage
%
%
\begin{figure}[th]
\centering
\includegraphics[width=9cm,angle=270]{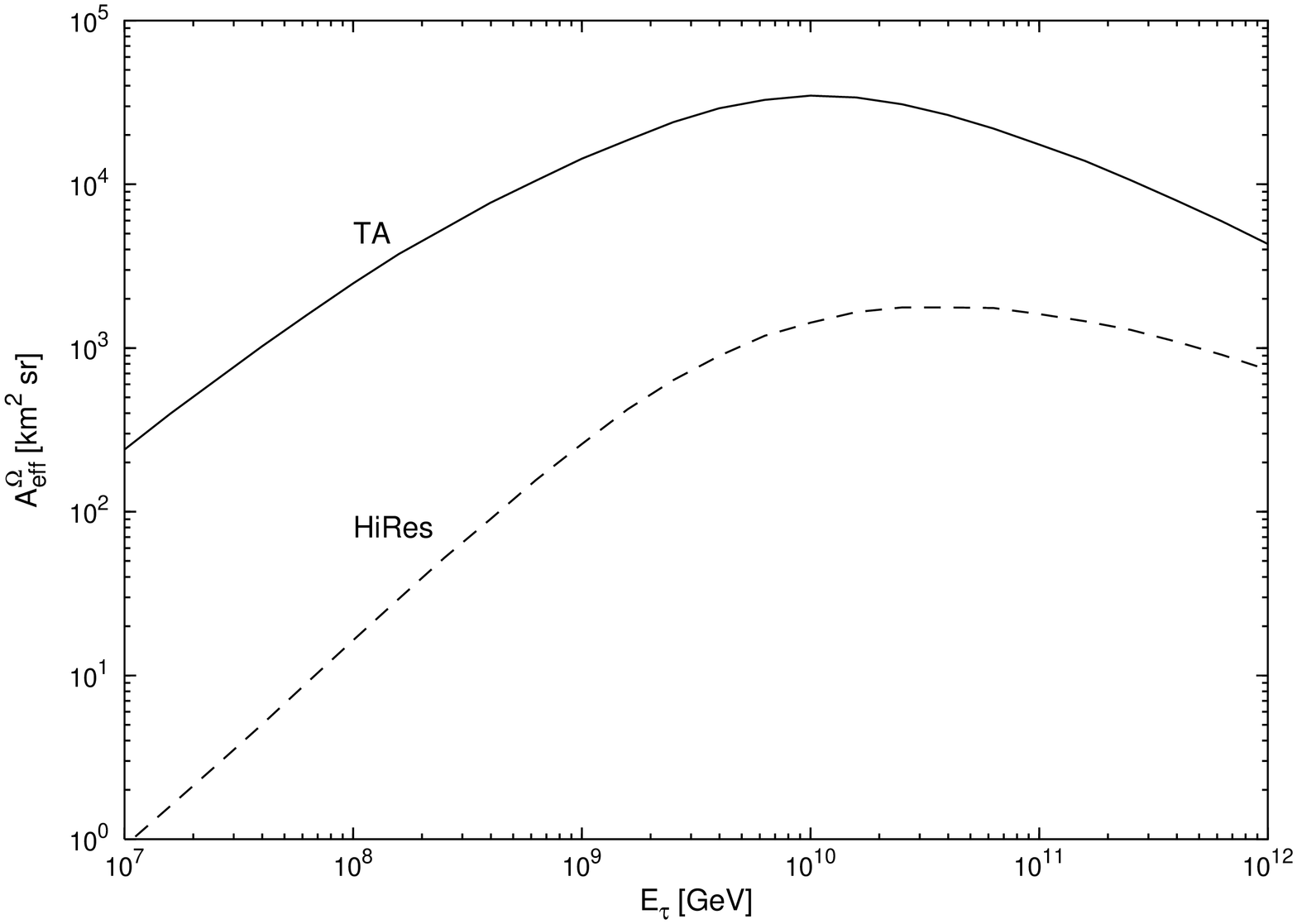}
\parbox{12cm}{
\medskip\footnotesize{Figure 9: Effective (geometric) aperture estimates [25] for the detection of Earth-skimming $\tau$ leptons through their most promising and dominant decays to electromagnetic showers for the Telescope Array (one station) and the HiRes detector. The TA aperture is exptected to be somewhat smaller [67] due to effects such as the detector response, light propagation in air, and night sky background.\newline}}
%
%
\includegraphics[width=9cm,angle=270]{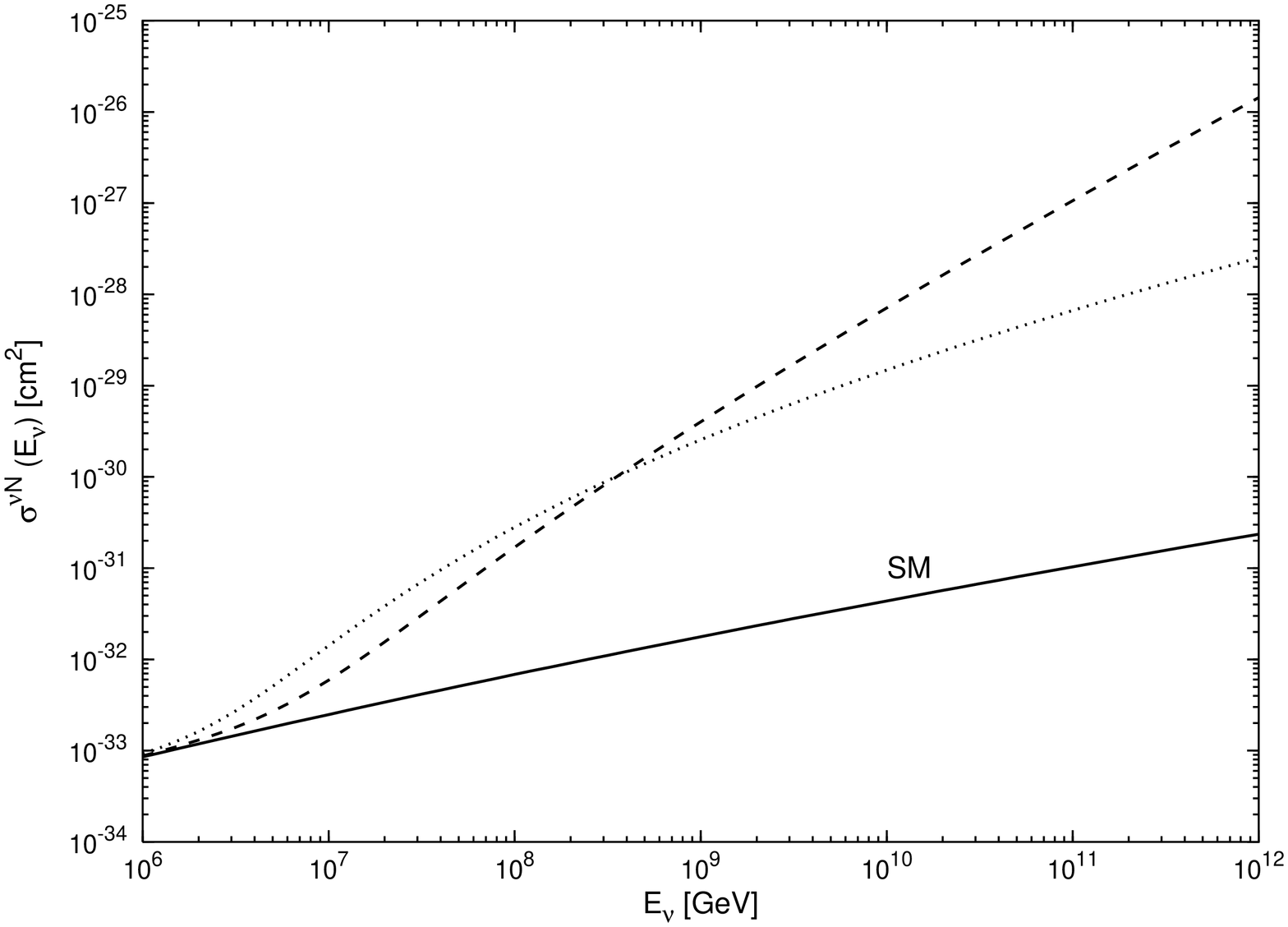}
\parbox{12cm}{
\medskip\footnotesize{Figure 10: The nominal total $\nu N$ cross section in the SM(CC+NC) compared to a scenario with large 'extra dimensions' using two different unitarity extrapolations between perturbative and nonperturbative regimes: the dottet line refers to Eq. (4.3) and the dashed one to Eq. (4.4).\newline}}
\end{figure}

\newpage

\begin{thebibliography}{54}
\bibitem{ref1}
R.~J.~Protheroe,
Nucl.\ Phys.\ Proc.\ Suppl.\  {\bf 77} (1999) 465.

\bibitem{ref2}
F.~Halzen,
Phys.\ Rept.\  {\bf 333} (2000) 349.

\bibitem{ref3}
D.~B.~Cline and F.~W.~Stecker,
astro-ph/0003459.

\bibitem{ref4}
A.~V.~Olinto,
astro-ph/0102077;

P.~Blasi,
astro-ph/0110401.

\bibitem{ref5}
G.~Sigl,
hep-ph/0109202;

P.~Bhattacharjee and G.~Sigl,
Phys.\ Rept.\  {\bf 327} (2000) 109
[astro-ph/9811011].

\bibitem{ref6}
K.~Greisen,
Phys.\ Rev.\ Lett.\  {\bf 16} (1966) 748;

G.~T.~Zatsepin and V.~A.~Kuzmin,
JETP Lett.\  {\bf 4} (1966) 78
[Pisma Zh.\ Eksp.\ Teor.\ Fiz.\  {\bf 4} (1966) 114].

\bibitem{ref7}
F.~W.~Stecker,
Astrophys.\ J.\  {\bf 228} (1979) 919.

\bibitem{ref8}
K.~Mannheim,
Astropart.\ Phys.\  {\bf 3} (1995) 295.

\bibitem{ref9}
F.~W.~Stecker and M.~H.~Salamon,
Space Sci.\ Rev.\  {\bf 75} (1996) 341
[astro-ph/9501064].

\bibitem{ref10}
F.~Halzen and E.~Zas,
Astrophys.\ J.\  {\bf 488} (1997) 669
[astro-ph/9702193].

\bibitem{ref11}
E.~Waxman and J.~N.~Bahcall,
Phys.\ Rev.\ D {\bf 59} (1999) 023002
[hep-ph/9807282].

\bibitem{ref12}
R.~J.~Protheroe and T.~Stanev,
Phys.\ Rev.\ Lett.\  {\bf 77} (1996) 3708;
[Erratum ibid.\  {\bf 78} (1997) 3420]
[astro-ph/9605036].

\bibitem{ref13}
V.~Berezinsky, M.~Kachelriess and A.~Vilenkin,
Phys.\ Rev.\ Lett.\  {\bf 79} (1997) 4302
[astro-ph/9708217];

M.~Birkel and S.~Sarkar,
Astropart.\ Phys.\  {\bf 9} (1998) 297
[hep-ph/9804285];

Z.~Fodor and S.~D.~Katz,
Phys.\ Rev.\ Lett.\  {\bf 86} (2001) 3224
[hep-ph/0008204];

S.~Sarkar and R.~Toldra,
Nucl.\ Phys.\ B {\bf 621} (2002) 495
[hep-ph/0108098];

C.~Barbot and M.~Drees,
Phys.\ Lett.\ B {\bf 533} (2002) 107
[hep-ph/0202072].

\bibitem{ref14}
G.~Sigl, S.~Lee, D.~N.~Schramm and P.~Coppi,
Phys.\ Lett.\ B {\bf 392} (1997) 129
[astro-ph/9610221].

\bibitem{ref15}
G.~Sigl, S.~Lee, P.~Bhattacharjee and S.~Yoshida,
Phys.\ Rev.\ D {\bf 59} (1999) 043504
[hep-ph/9809242].

\bibitem{ref16}
U.~F.~Wichoski, J.~H.~MacGibbon and R.~H.~Brandenberger,
Phys.\ Rev.\ D {\bf 65} (2002) 063005
[hep-ph/9805419].

\bibitem{ref17}
T.~J.~Weiler,
Phys.\ Rev.\ Lett.\  {\bf 49} (1982) 234;
Astropart.\ Phys.\  {\bf 11} (1999) 303
[hep-ph/9710431];
Astropart.\ Phys.\  {\bf 12} (2000) 379 (E);

D.~Fargion, B.~Mele and A.~Salis, Astrophys.\ J.\
{\bf 517} (1999) 725.

\bibitem{ref18}
S.~Yoshida, G.~Sigl and S.~Lee,
Phys.\ Rev.\ Lett.\  {\bf 81} (1998) 5505
[hep-ph/9808324].

\bibitem{ref19}
L.V.~Volkova, Yad.\ Fiz.\ {\bf 31} (1980) 1510,
[Sov.\ J.\ Nucl.\ Phys. {\bf 31} (1980) 784].

\bibitem{ref20}
R.~Gandhi, C.~Quigg, M.~H.~Reno and I.~Sarcevic,
Astropart.\ Phys.\  {\bf 5} (1996) 81
[hep-ph/9512364].

\bibitem{ref21}
R.~Gandhi, C.~Quigg, M.~H.~Reno and I.~Sarcevic,
Phys.\ Rev.\ D {\bf 58} (1998) 093009
[hep-ph/9807264].

\bibitem{ref22}
G.~M.~Frichter, D.~W.~McKay and J.~P.~Ralston,
Phys.\ Rev.\ Lett.\  {\bf 74} (1995) 1508
[Erratum ibid.\  {\bf 77} (1996) 4107]
[hep-ph/9409433].

\bibitem{ref23}
G.~C.~Hill,
Astropart.\ Phys.\  {\bf 6} (1997) 215
[astro-ph/9607140].

\bibitem{ref24}
C.~Hettlage and K.~Mannheim,
Proc. 2nd Workshop on Methodical Aspects of Underwater/Underice Neutrino Telescopes, Hamburg, 2001,
astro-ph/0202074.

\bibitem{ref25}
J.~L.~Feng, P.~Fisher, F.~Wilczek and T.~M.~Yu,
Phys.\ Rev.\ Lett.\  {\bf 88} (2002) 161102 and MIT-CTP-3122 [hep-ph/0105067].

\bibitem{ref26}
M.~Sasaki, Y. Asaoka and M.~Jobashi,
Univ.\ Tokyo ICRR-Report-484-2002-2 [astro-ph/0204167 v2].

\bibitem{ref27}
G.~Parente and E.~Zas, Proc. 7th Int. Symp. Neutrino Telescopes, Venice, Italy, 
Feb. 1996, p.499 [astro-ph/9606091].

\bibitem{ref28}
K.~S.~Capelle, J.~W.~Cronin, G.~Parente and E.~Zas,
Astropart.\ Phys.\  {\bf 8} (1998) 321
[astro-ph/9801313].

\bibitem{ref29}
M.~Gl\"uck, E.~Reya and A.~Vogt,
Z.\ Phys.\ C {\bf 48} (1990) 471.

\bibitem{ref30}
M.~Gl\"uck, E.~Reya and A.~Vogt,
Z.\ Phys.\ C {\bf 53} (1992) 127,
Phys.\ Lett.\ B {\bf 306} (1993) 391.

\bibitem{ref31}
M.~Gl\"uck, E.~Reya and A.~Vogt,
Z.\ Phys.\ C {\bf 67} (1995) 433.

\bibitem{ref32}
M.~Gl\"uck, E.~Reya and A.~Vogt,
Eur.\ Phys.\ J.\ C {\bf 5} (1998) 461
[hep-ph/9806404].

\bibitem{ref33}
I.~Abt {\it et al.}  [H1 Collaboration],
Nucl.\ Phys.\ B {\bf 407} (1993) 515;
T.~Ahmed {\it et al.},
Nucl.\ Phys.\ B {\bf 439} (1995) 471;
S.~Aid {\it et al.},
Phys.\ Lett.\ B {\bf 354} (1995) 494;

M.~Derrick {\it et al.}  [ZEUS Collaboration],
Phys.\ Lett.\ B {\bf 316} (1993) 412;
Z.\ Phys.\ C {\bf 65} (1995) 379;
Phys.\ Lett.\ B {\bf 345} (1995) 576.

\bibitem{ref34}
S.~Aid {\it et al.}  [H1 Collaboration],
Nucl.\ Phys.\ B {\bf 470} (1996) 3;
C.~Adloff {\it et al.},
Nucl.\ Phys.\ B {\bf 497} (1997) 3;

M.~Derrick {\it et al.}  [ZEUS Collaboration],
Z.\ Phys.\ C {\bf 69} (1996) 607;
Z.\ Phys.\ C {\bf 72} (1996) 399.

\bibitem{ref35}
J.~Kwiecinski, A.~D.~Martin and A.~M.~Stasto,
Phys.\ Rev.\ D {\bf 59} (1999) 093002
[astro-ph/9812262].

\bibitem{ref36}
M.~Gl\"uck, S.~Kretzer and E.~Reya,
Astropart.\ Phys.\  {\bf 11} (1999) 327
[astro-ph/9809273].

\bibitem{ref37}
H.~L.~Lai {\it et al.}, CTEQ3,
Phys.\ Rev.\ D {\bf 51} (1995) 4763
[hep-ph/9410404].

\bibitem{ref38}
H.~L.~Lai {\it et al.}, CTEQ4,
Phys.\ Rev.\ D {\bf 55} (1997) 1280
[hep-ph/9606399].

\bibitem{ref39}
E.~Aslanides {\it et al.}  [ANTARES Collaboration],
astro-ph/9907432.

\bibitem{ref40}
M.~Nagano and A.~A.~Watson,
Rev.\ Mod.\ Phys.\  {\bf 72} (2000) 689.

\bibitem{ref41}
X.~Bertou, M.~Boratav and A.~Letessier-Selvon,
Int.\ J.\ Mod.\ Phys.\ A {\bf 15} (2000) 2181
[astro-ph/0001516].

\bibitem{ref42}
D.~Fargion,
Astrophys.\ J.\  {\bf 570} (2002) 909
[astro-ph/0002453];
hep-ph/0111289 v2.

\bibitem{ref43}
X.~Bertou, P.~Billoir, O.~Deligny, C.~Lachaud and A.~Letessier-Selvon,
Astropart.\ Phys.\  {\bf 17} (2002) 183
[astro-ph/0104452].


\bibitem{ref44}
A.~Nicolaidis and A.~Taramopoulos,
Phys.\ Lett.\ B {\bf 386} (1996) 211
[hep-ph/9603382].

\bibitem{ref45}
V.~A.~Naumov and L.~Perrone,
Astropart.\ Phys.\  {\bf 10} (1999) 239
[hep-ph/9804301].

\bibitem{ref46}
D.~E.~Groom {\it et al.}  [Particle Data Group Collaboration],
Eur.\ Phys.\ J.\ C {\bf 15} (2000) 1, pp 152.

\bibitem{ref47}
A.~Dar, J.~J.~Lord and R.~J.~Wilkes,
Phys.\ Rev.\ D {\bf 33} (1986) 303.

\bibitem{ref48}
P.~Lipari and T.~Stanev,
Phys.\ Rev.\ D {\bf 44} (1991) 3543.

\bibitem{ref49}
A.~M.~Dziewonski and D.~L.~Anderson,
``Preliminary Reference Earth Model'',
Phys.\ Earth Planet.\ Interiors {\bf 25} (1981) 297.

\bibitem{ref50} 
A.~Biron {\it et al.},
''Proposal Upgrade of AMANDA-B towards AMANDA-II'', DESY-Zeuthen report (1997)

\bibitem{ref51}
M.~Leuthold,
``Ice Cube Configuration Studies'',
{\it Prepared for International Workshop on Simulations 
and Analysis Methods for Large Neutrino Telescopes, 
DESY--Zeuthen, Germany, 6-9 Jul 1998}

\bibitem{ref52}
S.~Fukuda {\it et al.}  [Super-Kamiokande Collaboration],
Phys.\ Rev.\ Lett.\  {\bf 85} (2000) 3999
[hep-ex/0009001];
Phys.\ Rev.\ Lett.\  {\bf 86} (2001) 5656
[hep-ex/0103033].

\bibitem{ref53}
J.~G.~Learned and S.~Pakvasa,
Astropart.\ Phys.\  {\bf 3} (1995) 267
[hep-ph/9405296].

\bibitem{ref54} 
H.~Athar, M.~Jezabek and O.~Yasuda,
Phys.\ Rev.\ D {\bf 62} (2000) 103007
[hep-ph/0005104];

L.~Bento, P.~Ker\"anen and J.~Maalampi,
Phys.\ Lett.\ B {\bf 476} (2000) 205
[hep-ph/9912240].

\bibitem{ref55}
J.~Alvarez-Muniz, F.~Halzen and D.~W.~Hooper,
Phys.\ Rev.\ D {\bf 62} (2000) 093015
[astro-ph/0006027].

\bibitem{ref56}
S.~I.~Dutta, M.~H.~Reno and I.~Sarcevic,
Phys.\ Rev.\ D {\bf 62} (2000) 123001
[hep-ph/0005310];
Phys.\ Rev.\ D {\bf 64} (2001) 113015
[hep-ph/0104275].

\bibitem{ref57}
S.~Bottai and S.~Giurgola,
Astropart.\ Phys.\  {\bf 18} (2003) 539
[astro-ph/0205325].

\bibitem{ref58}
P.~Jain, D.~W.~McKay, S.~Panda and J.~P.~Ralston,
Phys.\ Lett.\ B {\bf 484} (2000) 267
[hep-ph/0001031].

\bibitem{ref59}
M.~Kachelriess and M.~Pl\"umacher,
Phys.\ Rev.\ D {\bf 62} (2000) 103006
[astro-ph/0005309].

\bibitem{ref60}
F.~Cornet, J.~I.~Illana and M.~Masip,
Phys.\ Rev.\ Lett.\  {\bf 86} (2001) 4235
[hep-ph/0102065].

\bibitem{ref61}
A.~Donnachie and P.~V.~Landshoff,
Phys.\ Lett.\ B {\bf 437} (1998) 408
[hep-ph/9806344];
Phys.\ Lett.\ B {\bf 518} (2001) 63
[hep-ph/0105088].

\bibitem{ref62}
V.~S.~Berezinsky, A.~Z.~Gazizov and S.~I.~Yanush,
Phys.\ Rev.\ D {\bf 65} (2002) 093003
[astro-ph/0105368].

\bibitem{ref63}
Auger Collaboration,
``The Pierre Auger Project Design Report'',
FERMILAB-PUB-96-024, Feb. 1997, http://www.auger.org.

\bibitem {ref64} 
TA Collaboration,
''The Telescope Array Project Design Report'', July 2000, http://www-ta.icrr.u-tokyo.ac.jp

\bibitem{ref65} 
P.~Billoir, 
Pierre Auger project note GAP-97-049,
L.P.N.H.E.\ Paris VI--VII
(http://www.auger.org/admin-cgi-bin/woda/gap\_notes.pl)


\bibitem{ref66}
A.~Kusenko and T.~J.~Weiler,
Phys.\ Rev.\ Lett.\  {\bf 88} (2002) 161101
[hep-ph/0106071].

\bibitem{ref67} 
Y.~Asaoka, M.~Jobashi and M.~Sasaki, private communication.

\bibitem{ref68}
Particle Data Group, J.~Hewett and J.~March-Russell, 
Phys.\ Rev.\ D {\bf 66} (2002) 010001-945, and references therein.

\bibitem{ref69}
T.~Han, J.~D.~Lykken and R.-J.~Zhang,
Phys.\ Rev.\ D {\bf 59} (1999) 105006
[hep-ph/9811350];

G.~F.~Giudice, R.~Rattazzi and J.~D.~Wells,
Nucl.\ Phys.\ B {\bf 544} (1999) 3
[hep-ph/9811291].

\bibitem{ref70}
P.~Mathews, S.~Raychaudhuri and K.~Sridhar,
Phys.\ Lett.\ B {\bf 455} (1999) 115
[hep-ph/9812486].

\bibitem{ref71}
S.~Nussinov and R.~Shrock,
Phys.\ Rev.\ D {\bf 59} (1999) 105002
[hep-ph/9811323].

\bibitem{ref72} 
C.~Tyler, A.V.~Olinto and G.~Sigl,
Phys.\ Rev.\ {\bf D63} (2001) 055001
[hep-ph/0002257].

\bibitem{ref73} 
L.A.~Anchordoqui, J.L.~Feng, H.~Goldberg and
A.D.~Shapere, 
Phys.\ Rev.\ {\bf D66} (2002) 103002
[hep-ph/0207139 v2].

\bibitem{ref74} 
O.E.~Kalashev {\it et al.}, 
Phys.\ Rev.\ {\bf D66} (2002) 063004.

\bibitem{ref75} 
J.~Alvarez-Mu\~niz, F.~Halzen, T.~Han and D.~Hopper,
Phys.\ Rev.\ Lett.\ {\bf 88} (2002) 021301.

\bibitem{ref76}
S.~Yoshida {\it et al.},
Astropart.\ Phys.\  {\bf 3} (1995) 105; www.icrr.u-tokyo.ac.jp/as/project/
agasa.html

\bibitem{ref77}
D.~J.~Bird {\it et al.}  [HIRES Collaboration],
Astrophys.\ J.\  {\bf 424} (1994) 491; http://hires.physics.utha.edu;
T.~Abu-Zayyad {\it et al.}  [High Resolution Fly's Eye Collaboration],
astro-ph/0208243 v2.

\bibitem{ref78}
L.~A.~Anchordoqui, J.~L.~Feng, H.~Goldberg and A.~D.~Shapere,
Phys.\ Rev.\ D {\bf 65} (2002) 124027
[hep-ph/0112247].
\end{thebibliography}
\end{document}